\documentclass[letterpaper,superscriptaddress,aps,pra,twocolumn,showpacs,floatfix,longbibliography]{revtex4-2} % {{{
\usepackage{natbib}

\usepackage{hyperref}
\usepackage{graphicx,helvet}
\usepackage{color}
\usepackage{bm}                      % added by TG
\usepackage{tikz}
\usepackage{amsfonts}
\usepackage{lipsum}
\definecolor{mygray}{gray}{0.4}
\definecolor{light-blue}{rgb}{0.38,0.58,0.69}
\definecolor{myred}{rgb}{0.83,0.34,0.37}
\definecolor{mygreen}{rgb}{ .59,.74,.61}
\definecolor{miviol}{rgb}{.57,.11,.81} %%{0.55,0.48,0.80}
\usepackage{amsmath}%
\usepackage{MnSymbol}%
\usepackage{wasysym}%
 \usepackage{soul}
\usetikzlibrary{decorations.shapes}

\newcommand{\kb}{k_{\rm B}}
\newcommand{\beqa}{\begin{eqnarray}}
\newcommand{\eeqa}{\end{eqnarray}}
\newcommand{\beq}{\begin{equation}}
\newcommand{\eeq}{\end{equation}}

\usepackage{amsmath}

\newcommand{\mcC}{\mathcal{C}}

\newcommand{\tr}{\mathop{\mathrm{Tr}}\nolimits}

\def\rv{\mathbf{r}}
\def\rvp{\mathbf{r}^{\prime}}

\def\rb{\mathbf{\bar r}}
\def\pv{\mathbf{p}}

\def\pb{\mathbf{\bar p}}
\def\vxi{\boldsymbol{\xi}} 
\def\vchi{\boldsymbol{\chi}}
\def\ve{\varepsilon}
\def\d{{\rm d}}

\newcommand{\ifimar}{Instituto de Investigaciones F\'isicas de Mar del Plata (IFIMAR), CONICET-UNMdP,  Mar del Plata,  Argentina}
\newcommand{\conicet}{Consejo Nacional de Investigaciones Cient\'ificas y Tecnol\'ogicas (CONICET), Argentina}
\newcommand{\uba}{Departamento de F\'isica ``J. J. Giambiagi'' and IFIBA, FCEyN, Universidad de Buenos Aires, 1428 Buenos Aires, Argentina}
\newcommand{\ipcms}{\mbox{Universit\'e de Strasbourg, CNRS, Institut de Physique et Chimie des Mat\'eriaux 
de Strasbourg,} UMR 7504, F-67000 Strasbourg, France}

\newcommand{\hX}{\hat{X}}
\newcommand{\hP}{\hat{P}}
\newcommand{\hQ}{\hat{Q}}
\newcommand{\hU}{\hat{U}}
\newcommand{\hV}{\hat{V}}
\newcommand{\hW}{\hat{W}}

\newcommand{\hO}{\hat{O}}

\newcommand{\im}{{\rm i}}
\newcommand{\tE}{t_{\mathrm E}}
\def\H{\hat{H}}
\begin{document}
\title{Out-of-time-order correlators and quantum chaos}
\author{Ignacio Garc\'ia-Mata} \affiliation{\ifimar} \affiliation{\conicet}
\author{Rodolfo A. Jalabert} \affiliation{\ipcms} 
\author{Diego A. Wisniacki} \affiliation{\uba}
% \email{@gmail.com}

\date{September 16, 2022}
\begin{abstract}
Quantum Chaos has originally emerged as the field which studies how the properties of classical chaotic systems arise in their quantum counterparts. The growing interest in quantum many-body systems, with no obvious classical meaning has led to consider time-dependent quantities that can help to characterize and redefine Quantum Chaos.  This article reviews the prominent role that the out of time ordered correlator (OTOC) plays to achieve such goal. 
\end{abstract}

\maketitle

\section{Introduction}
\subsection{Basic concepts}
The field of Quantum Chaos, developed from the early eighties and addressing the quantum manifestations of an underlying classically chaotic dynamics, has primarily been concerned with the time-independent one-body case. Different tools have been developed to bridge the gap between classical and quantum mechanics \cite{gutzwiller2007}, linking properties that have a correspondence in the two realms, like for instance, the chaotic nature of the classical dynamics and the level statistics of the quantum states \cite{bohigas,ullmo2016}. The generalization of such a scheme to setups with non-chaotic dynamics (integrable or mixed), time and/or temperature dependent cases, as well as many-body systems with or without classical analogue, presents an important challenge, which is typically encountered when approaching experimentally relevant situations \cite{raizen2011,jalabert2016}.

The generalization of quantum chaos studies with regards to time-evolution problems faces an important obstacle even in the case of a time-independent Hamiltonian; in order to monitor the behavior of physical quantities, we need to characterize the evolution of operators, taking place in vector spaces with greater complexity than the usual Hilbert space of quantum states. It is then difficult to establish general results, and therefore new tools are needed to characterize the properties of operator dynamics according to the characteristics of the quantum system under study. This difficulty, together with the fact that until recently the available experimental techniques were essentially developed for the stationary case, are probably responsible for quantum dynamics studies lagging behind those of time-independent problems. Fortunately, the presently available resources to follow and control the time-evolution of complex quantum systems have changed the states of affairs, contributing to a sustained interest in the subject of quantum dynamics and to the development of new theoretical tools.  

Early studies of stochasticity in the dynamics of complex quantum systems considered the two-point correlation function of Heisenberg operators at different times, establishing a departure  from the corresponding classical correlator after a relatively short time \cite{shepelyanskii1981,shepelyansky1983}. In addition, the sensitivity with respect to an imperfect time-reversal in complex quantum systems has been analyzed from a quantum chaos perspective in the framework of the Loschmidt echo. It was there shown that, in classically chaotic systems, the rate of fidelity-loss could become independent, within a limited time-window, of the degree of imperfection in the time-reversal protocol \cite{Jalabert2001,DiegoScholar}.  

In the progression from one-body to many-body systems, one needs to consider that 
the increasing complexity of the dynamics stems from the number of particles as well as from the nature of the interactions \cite{ullmo2008}. The non-interacting many-body systems present the same degree of complexity in their dynamics than that of the corresponding one-body case, but their analysis requires the incorporation of restrictions arising from the particle statistics (i.e. fermionic or bosonic) \cite{bohigas1999,leboeuf2005,hummel2013}. Weakly-interacting many-body systems (like a particle in contact with an environmental bath or a long-lived quasi-particle excitation in a Fermi liquid) exhibit no more complexity than of the one-body case, up to long-time effects such as thermalization with the bath, decoherence, or relaxation. The quantum chaos studies must take into account the time and length scales where the one-body dynamics is relevant (i.e. phase-coherence length, inelastic mean-free-path) \cite{jalabert1990,jalabert2016}.

The next level of complexity arises when addressing strongly interacting many-body systems, where quasi-particles are not well-defined. A precursor of quantum chaos studies in strongly interacting many-body systems was provided by Wigner's landmark observation that the level statistics of compound nucleus agreed with that of the Gaussian orthogonal ensemble (GOE) predicted by random-matrix-theory \cite{porter}. Taking into account the correlations between the elements of the Hamiltonian matrix by the Pauli principle and a two-body random interaction, lead to the Bohigas-Flores-French-Wong two-body random ensemble (TBRE) \cite{French1970,Bohiga1971,BrodyRMP}, that resulted in a more realistic eigenvalue distribution than the semi-circle law of the GOE.
Numerical studies of level statistics concerning wide sectors of the spectrum corresponding to non-integrable condensed matter many-body systems also yielded the correlations of random-matrix theory \cite{poilblanc1993}. Outside the studies of level statistics of highly-excited states in many-body systems, it is difficult to establish general results, specially when addressing the non-stationary case or treating systems without classical analogue. 

Within this context, the out-of-time-order correlator (OTOC), describing the averaged evolution of quantum Heisenberg operators at different times,
appears as a valuable tool for generalizing quantum chaos studies beyond the time-independent, one-body case. Introduced by Larkin and Ovchinnikov in 1969 in order to discuss the applicability of semiclassical approaches to superconductivity \cite{larkin1969quasiclassical}, the OTOC started to receive considerable attention much latter on, following the remark about its relevance for studies relating black-hole horizon geometry and chaos \cite{shenker2014black,shenker2014multiple,shenker2015stringy}. 

Black-holes were shown to exhibit the early exponential time-growth of the OTOC expected in classically chaotic systems, leading to the popular terminologies of ``scrambling'', ``quantum butterfly effect'', and ``quantum Lyapunov exponent''. Such a behavior enabled to place black-holes in the class of fast scramblers \cite{sekino2008fast}, where the information is rapidly spread. Moreover,  Maldacena, {\em et al.} \cite{maldacena2016} conjectured in 2016 the existence of an upper bound to the growth-rate of the OTOC, essentially given by the temperature of the system. 

Studies in other realms of Theoretical Physics followed, and for the strongly correlated fermionic systems described by the Sachdev-Ye-Kitaev model, the previous bound was also shown to saturate, placing them also in the category of fast scramblers \cite{kitaev2015simple}. The relevance of the OTOC in such diverse fields is at the origin of the substantial interest that it has received in recent times \cite{swingle2018,swingle2022}. 

The purpose of this article is to present the general properties of the OTOC, focusing on the relation between scrambling -- as measured by the OTOC --  and quantum chaos. We review the recent key developments in the domain, and analyze how the main features of quantum chaos are contained in the physics and the behavior of the OTOC. Moreover, we describe the OTOC as an established invaluable tool to comprehend and characterize complex systems, especially when undergoing a transition from regular to chaotic regime and when no classical analogue exists to guide our intuition.

\subsection{Definition}
\label{subsec:definition}
%\noindent

The OTOC is a time-dependent function defined by an averaged double-commutator as
\begin{equation}
\label{otocdef1}
{\cal C}_{\hV\hW}(t)=\left \langle [\hW_t,\hV]^{\dagger}[\hW_t,\hV]\right \rangle \,  .
\end{equation}

\begin{itemize}
\item $\hV$ and $\hW$ are operators in a Hilbert space $\cal H$.
\item The sub-index $t$ stands for the time-evolution of the operator $\hO$ under the time-independent Hamiltonian $\H$ in the Heisenberg representation,  {\it i.e.}  $\hO_t=\hU_t^{\dagger}\, \hO \, \hU_t$ with $\hU_t=e^{-\im \H t/\hbar}$ the unitary  time-evolution operator.
\item $ \langle \, \cdot\,\rangle = \tr \left\{ \rho \, \cdot\, \right\}$ represents the thermal average in the canonical ensemble of constant number of particles.
\item $\rho= Z^{-1} \, e^{-\beta \H}$ is the thermal density matrix (or density operator),  $Z$ is the canonical partition function,  $\beta=(\kb T)^{-1}$, with $T$ the temperature and $\kb$ the Boltzmann constant. 
\end{itemize}

The definition \eqref{otocdef1} holds for an arbitrary pair of operators.  However,  with the goal of characterizing physical observables and in the search of possible universal behavior, the definition of the OTOC is typically restricted to the case in which operators $\hV$ and $\hW$ are Hermitian,  where
\begin{equation}
\label{otocHerm}
{\cal C}_{\hV\hW}^{\rm H}(t)=- \left \langle [\hW_t,\hV]^{2}\right \rangle \,  .
\end{equation}
Given their physical relevance and their connection with Hermitian operators, the choice of a pair of unitary operators is also of interest. In addition, the mixed case of one operator taken to be Hermitian and the other unitary is sometimes considered. Unless explicitly specified otherwise, this article addresses the case of Hermitian operators.

Developing \eqref{otocdef1},  the OTOC in the generic case can be expressed as 
\begin{equation}
{\cal C}_{\hV\hW}(t) = {\cal D}_{\hV\hW}(t) + {\cal I}_{\hV\hW}(t) - 2 \,  {\rm Re} \left\{ {\cal F}_{\hV\hW}(t)\right\} \, .
\label{otocdef2}
\end{equation} 

\begin{itemize}
\item ${\cal D}_{\hV\hW}(t)= \left \langle \hV^{\dagger} \left( \hW^{\dagger} \hW \right)_t \hV \right \rangle$.
\item ${\cal I}_{\hV\hW}(t)= \left \langle \hW_t^{\dagger} \,  \hV^{\dagger} \, \hV \,  \hW_t \right \rangle$.
\item ${\cal F}_{\hV\hW}(t)= \left \langle \hW_t^{\dagger} \,   \hV^{\dagger} \,  \hW_t \,  \hV \right \rangle$.
\end{itemize}

In the Hermitian case, the choice $\hV = \hW$, as well as an infinite temperature, results in ${\cal I}_{\hV\hW}(t) = {\cal D}_{\hV\hW}(t)$. 

In the unitary case,  ${\cal D}_{\hV\hW}(t) = {\cal I}_{\hV\hW}(t) = 1$, and then 
\begin{equation}
\label{otocUni}
{\cal C}_{\hV\hW}^{\rm uni}(t)= 2\left(1- {\rm Re} \left\{ {\cal F}_{\hV\hW}(t)\right\} \right) \,  .
\end{equation}

In general, the 3-point functions ${\cal D}_{\hV\hW}(t)$ and ${\cal I}_{\hV\hW}(t)$, and the 4-point function ${\cal F}_{\hV\hW}(t)$, are time-correlators of different complexity. While ${\cal D}_{\hV\hW}(t)$ counts only two evolution operators in its definition, ${\cal I}_{\hV\hW}(t)$ and ${\cal F}_{\hV\hW}(t)$ depend on four evolution operators. The simplest correlator is thus ${\cal D}_{\hV\hW}(t)$, which can be expressed as a time-ordered product, while ${\cal F}_{\hV\hW}(t)$ is a genuine out-of-time-order product expected to exhibit a non-trivial time behavior. This is why ${\cal F}_{\hV\hW}(t)$ is sometimes referred to as OTOC. But we will avoid this ambiguous nomenclature. The behavior of ${\cal I}_{\hV\hW}(t)$ depends on the system under consideration, and might be related in some cases to that of ${\cal F}_{\hV\hW}(t)$ \cite{Ueda2018,JGMW2018}.

The choice of $\hV = \hW$ trivially implies that ${\cal C}_{\hV\hV}(0) = 0$.  The case of a pair of canonical conjugated (or complementary) operators fulfilling a special equal-time commutation relation (like the $x$-component of the position and the momentum, $\hat X$ and $\hat P_X$, respectively) constitutes another choice of interest characterized by a non-vanishing initial value of the OTOC. 

It is common to assume that the operators $\hV$ and $\hW$ have local support. That is,  they are associated with a localized subsystem of ${\cal H}$.  The operator $\hat X$ in a system with classical analogue and the spin component ${\hat S}_{i}^{\mu}$ within a spin-chain ($i=0,1, ... , L-1$; $\mu = x,y,z$) constitute  examples of fully localized operators.  When the operators $\hW$ and $ \hV$ are defined at specific positions (like for instance the sites $i=0$ and $i=l$ in a spin chain, respectively) the OTOC can be expressed as function of time and distance ($t$ and $l$),  allowing to address the space-time propagation of an initial excitation (as discussed in the next chapter). 

Sometimes,  instead of \eqref{otocdef1},  a regularization is adopted for the definition of the OTOC,  by smearing the thermal distribution between the two commutators, using
\begin{equation}
\label{otocreg}
{\cal C}_{\hV\hW}^{\rm reg}(t)=  \tr \left\{ \rho^{1/2} \,  [\hW_t,\hV]^{\dagger} \, \rho^{1/2} \, [\hW_t,\hV] \right\} \,  .
\end{equation}
Such a regularization  is introduced for calculation purposes,  assuming that the long-time properties of the OTOC are independent of it \cite{chowdhury2017}. However,  such an expectation has been found not to hold in general \cite{liao2018nonlinear,romero2019,sahu2020}. Regularization other than that of Eq.~\eqref{otocreg} have been proposed \cite{tsuji2018bound}, and used in the study of the general properties of the OTOC.

\subsection{Physical interpretation}
The OTOC defined by Eq.~\eqref{otocdef1} provides a valuable tool for the generalization of quantum chaos discussed in the introductory chapter. It is manifestly time and temperature dependent, and can be used for a wide class of physical systems, ranging from one-body to many-body, as well as from completely integrable to completely chaotic, and moreover, having or not a classical analogue.

\begin{figure}
    \centering
    \includegraphics[width=0.95\linewidth]{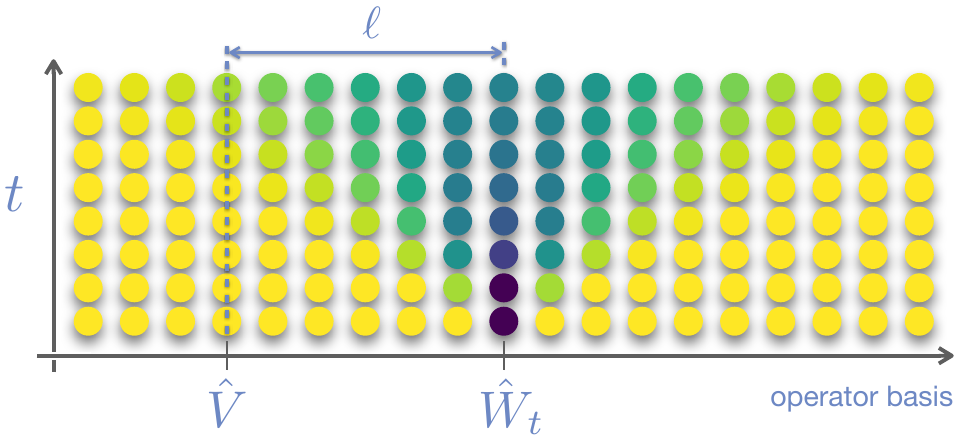} %{figs/fig_operator_spread.pdf}  %%{fig1_alt_light.pdf}
    %{fig1-2.png}
    \caption{Sketch illustrating how an operator $\hW$, localized at the central site of a one-dimensional system (like for instance a spin-chain), spreads among an operator basis, according to the time-evolution of the Heisenberg operator $\hW_t$. A second operator $\hV$, associated with a site at a distance $l$ from the center, is initially unaffected by the perturbation (yellow dots), but enters at latter times in the propagation cone of $\hW_t$ (green dots).}
    \label{Figsread}
\end{figure}

The growth of the OTOC is associated with the spread of quantum information,  a process commonly referred to as information scrambling \cite{shenker2015stringy}. Such a spread can be quantified in the operator space by taking $\hW$ and $\hV$ as operators that locally act in space-like separated regions (and thus $[\hW,\hV]=0$).  Following Ref.~\cite{swingle2018},  Fig.~\ref{Figsread} sketches the case of a one-dimensional system, with $\hW$ localized at the middle of the chain, and $\hV$ located $l$ sites away to the left. The Baker-Campbell-Hausdorff formula for the expansion of $\hW_t$ yields the nested commutator expression
\begin{equation}
\label{eqBCH}
\hW_t=\sum_{k=0}^{\infty} \frac{(i t)^{k}}{k !} \underbrace{[\H, \ldots[\H}_{k}, \hW] \ldots]
\end{equation}
If the Hamiltonian contains local $n$-body interactions (for instance,  $n=2$ for the case of a nearest-neighbor coupling in a spin chain),  from Eq~(\ref{eqBCH}) it can be inferred that as time increases so does the number of sites involved,  as illustrated by the propagation cone depicted in Fig.~\ref{Figsread}.  Initially,  only the central site is "on" (indicated by the darkest color), and the rest of the system is unaffected (yellow sites). As the time-evolution takes place according to $\H$,  other sites are involved ({\it i.e.} higher-order terms in the expansion \eqref{eqBCH} become relevant),  and in particular the site associated with the operator $\hV$.  Such an effect can be probed by the non-vanishing commutator of $\hV$ with $\hW_t$,  and thus by the OTOC. The early approach to scrambling is expected to depend on the variable $t-l/v_{\rm B}$, typically in an exponential way, ensuring that ${\cal C}_{\hV\hW}(t)$ is very small when $t \ll l/v_{\rm B}$, and thus $\hV$ is outside the propagation cone of $\hW_t$. The state-dependent butterfly velocity $v_{\rm B}$ sets the slope of the propagating cone in Fig.~\ref{Figsread}, and it is limited by the Lieb-Robinson bound \cite{LiebRobinson72} for the speed of correlations spreading in non-relativistic quantum systems \cite{roberts_swingle_2016,Khemani2018}. The image of Fig.~\ref{Figsread} appears as particularly simple due to the choice of operators localized on the sites of a one-dimensional system. Moreover, as will be seen in the sequel, the dependence on $t-l/v_{\rm B}$ might take other forms than a simple exponential \cite{Swingle2019,DasSarma2019,swingle2020}.

An alternative view of characterizing the time-evolution of Heisenberg operators follows from the observation that the overlap of the states $\hV \hW_t  |\psi \rangle$ and $\hW_t\ \hV |\psi \rangle$, originated from the same initial state $|\psi \rangle$ subjected to the action of $\hV$ and $\hW_t$ in different order, yields the zero-temperature ${\cal F}_{\hV\hW}(t)$ if $|\psi \rangle$ is chosen as the ground state, or the infinite-temperature ${\cal F}_{\hV\hW}(t)$ if the trace over the states $|\psi \rangle$ is taken. Therefore, ${\cal F}_{\hV\hW}(t)$ represents a fidelity measuring the non-commutativity between the application of a local operation (given by $\hV$) and a time-evolved operator ($\hW_t$). Only in the trivial case $[\hW_t,\hV]=0$ for all $t$, would  ${\cal F}_{\hV\hW}(t)=1$ and ${\cal C}_{\hV\hW}(t)=0$ be valid.

The sequence of the operators $\hU_t$ and $\hU_t^{\dagger}$ appearing in the expressions of ${\cal C}_{\hV\hW}(t)$ and ${\cal F}_{\hV\hW}(t)$ can be interpreted as the alternation of forward and backward time-evolution, resulting in a structure reminiscent of the Loschmidt echo. The latter is a measure of the revival occurring when an imperfect rewinding of time is applied to a complex quantum system, and allows at the same time to quantify the sensitivity of the quantum evolution to perturbations through \cite{goirin_PhysRep2006,Jacquod2009,DiegoScholar,goussev2016}
\begin{equation}
M(t)=\left|\langle \psi |e^{\im (\H+ {\hat \Sigma}) t/\hbar} \, e^{-\im \H t/\hbar}|\psi \rangle \right|^2 \, ,
\end{equation}
\begin{itemize}
\item $|\psi \rangle$ is an arbitrary initial state, typically chosen to be a coherent state.
\item The operator ${\hat \Sigma}$ characterizes the quality of the time-reversal operation.
\end{itemize}

The links between the OTOC and the Loschmidt echo have been explored in particular systems, collating the similarities and differences of their respective time-behavior \cite{sreeram2021,yan_zurek2020,bhattacharyya2022towards}.
The important difference between the OTOC and the Loschmidt echo resides in the imperfect time reversal that defines the latter, introducing a random component or the interaction with extra degrees of freedom (i.e. an environment).
A formal connection between ${\cal F}_{\hV\hW}(t)$ and $M(t)$ can be established for particular operator choices. For instance, taking $\hV$ as a density operator associated with a pure state ({\it i.e.} $\hV = |\psi \rangle \langle \psi |$) and $\hW$ corresponding to the unitary evolution determined by a weak quasi-instantaneous quench on a short time-interval $\delta \tau$, while working in the infinite-temperature limit, results in an out-of-time-order product
\begin{equation}
{\cal F}_{\hV\hW}(t)=
\left|\langle \psi |e^{\im \H t/\hbar} \, e^{-\im {\hat \Sigma} \delta \tau/\hbar} \, e^{-\im \H t/\hbar}|\psi \rangle \right|^2 \, ,
\end{equation}
which coincides with the Loschmidt echo for the case of a small kick perturbation acting at time $t$   \cite{kurchan2016,schmitt2018,schmitt2019,Sanchez2021}. Outside this particular choice, the link between the two quantities requires an averaging procedure, and relies on the independence of the Loschmidt echo on ${\hat \Sigma}$, occurring in certain regimes
\cite{Jalabert2001}. A statistical connection has been established between ${\cal F}_{\hV\hW}(t)$ and $M(t)$, when the first is averaged over the defining unitary operators $\hV$ and $\hW$ (with the Haar measure), while the second is thermally averaged \cite{yan_zurek2020}. This relation was further extended for high-order OTOCs, beyond 4-point functions, and multi-fold Loschmidt echoes (with more than one forward-backward loop) \cite{bhattacharyya2022towards}. The square of the magnetization fidelity with small random rotations at the moment of the time-reversal can be expressed as a correlation function of the OTOC type. Early works in chains of spin $1/2$ or with large spin using particular protocols of time reversal were shown to exhibit exponential sensitivity in the classical case in contrast with non-exponential behavior in the quantum case \cite{fine2014absence,elsayed2015sensitivity}.

While the above discussion presented a valuable insight on how the time-growth of the OTOC characterizes the evolution of quantum dynamics, from a purely quantum mechanics point of view, it not obvious to anticipate about the specific behavior of the OTOC as a function of time, nor the existence of universal features that make the OTOC an interesting physical quantity. Fortunately, the quasi-classical limit of $\hbar \rightarrow 0$ is quite appropriate in order to address the previous issues, as acknowledged at the time of introducing the OTOC \cite{larkin1969quasiclassical}. Given the well-known relation between the commutator of two operators and the Poisson brackets of the corresponding classical symbols (considered as phase-space functions) 
\begin{equation}
    \lim _{\hbar \rightarrow 0} \frac{1}{i \hbar}[\hat{W}, \hat{V}]=\{W, V\} \, ,
\end{equation}
while making the choice $\hW=\hat{X}$ and $\hV=\hat{P}_X$, the quasi-classical limit of the commutator $[\hat{X}_t,\hat{P}_X]$ can be expressed as
\begin{equation}
    \im \hbar \, \{X(t), P_X(0)\}=i \hbar \, \frac{\partial X(t)}{\partial X(0)} \, .
\end{equation}
If the classical dynamics is fully chaotic, the exponential sensitivity with respect to the initial condition translates into $\partial X(t)/\partial X(0) \sim \exp (\lambda t)$, where $\lambda$ is the Lyapunov exponent. Therefore, the quasi-classical limit of the OTOC is given by 
\begin{equation}
\label{eq:qcOTOC}
    \mathcal{C}_{\hat P_X \hat X}^{\rm qc}(t) \sim \hbar^{2} \exp (2 \lambda t) \, .
\end{equation} 

 This relationship is particularly interesting, since it establishes the signature of the Lyapunov exponent, which is a purely classical quantity, in the OTOC, which is a genuine a quantum object. Such an unusual link also appears for the Loschmidt echo \cite{Jalabert2001}, underlying the above mentioned interest in determining connections between the later and the OTOC.

\section{Time regimes}
\label{sec:time_reg}
While the quasi-classical form \eqref{eq:qcOTOC} of the OTOC points to the emergence of classical features in quantum systems, it is clear that such kind of universal behavior can only appear in a well-defined class of systems, and in a particular time-window. It is therefore of fundamental importance to describe the different time-regimes that can be observed in the evolution of the OTOC. The 
intertwined analysis concerning time-windows and system features  will be structured in the sequel by first discussing, in the present chapter, the various time-regimes encountered in the OTOC evolution, and then, in the ensuing chapter about OTOC and quantum chaos, presenting a comprehensive survey about the relevance of the system characteristics on the OTOC behavior.  

\begin{figure}
    \centering
    \includegraphics[width=0.95\linewidth]{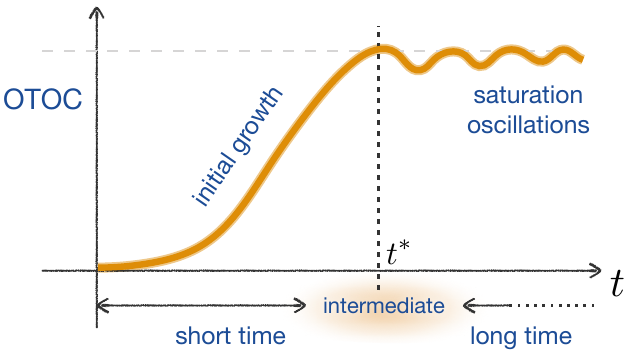} %%{figs/esquema_time_regime.pdf}  
     \caption{ Schematic picture for the time-evolution of the OTOC. The initial growth is expected to be exponential if the system is chaotic, but may take different forms in other cases. After the scrambling time $t^*$ the OTOC settles donw, showing oscillations around a constant value. In the chaotic case the oscillations are strongly suppressed and the OTOC evolves towards an approximately constant value. Such an approach is governed by the largest (classical) Ruelle-Pollicott resonance.}
    \label{Figtimescheme}
\end{figure}

According to what can be observed from the time-evolution in different systems across many previous works, 
the OTOC description can in general be divided in three time-regimes, schematically indicated in Fig.~\ref{Figtimescheme}. 

The short-time regime corresponds to the initial growth, where operator $\hW_t$ spreads, driven by the Hamiltonian dynamics.  

At some time $t^*$, called scrambling time, the OTOC growth stops.  We dub  intermediate time-regime the transition interval around $t^*$, where the OTOC settles down. As will be later discussed, the intermediate time-regime is characterized by the decay of ${\cal F}_{\hV\hW}$. 

Later on, in the long-time regime the mean value of the OTOC remains approximately constant, with eventual oscillations superposed, depending on the system under consideration. 

The remaining of this section describes the main features encountered in each of these time-regimes, using the example of quantum maps as a guide, and commenting on related behavior found in other physical systems. 

%
%%%%%%%%%%%%%%%%%%%%%%%%%%%%%%%%%%%
%%
\subsection{Short times}
As the operator $\hW_t $ evolves according to \eqref{eqBCH}, first there appears a perturbative regime where a power-law growth of the OTOC is expected to hold. After this trivial initial takeoff of the OTOC, but for sufficiently short times, that do not reach the scrambling time, the monotonous growth continues. It is for these short times that the OTOC growth has been more thoroughly studied, and predictions like that of Eq.~\ref{eq:qcOTOC} have been put forward. 

Quantum one-body systems with a  strongly chaotic classical counterpart, like quantum billiards or quantum maps, are natural candidates to exhibit such a behavior, and this possibility has been investigated for different setups (see e.g.  \cite{Rozenbaum2017,OTOC_gato_PRL,JGMW2018,lakshminarayan2019out,chavez2019quantum,rozenbaum2019} to name a few). While there has been some debate  about the correspondence between the exponent $\Lambda$ obtained as the rate of the OTOC growth and the corresponding Lyapunov exponent $\lambda$ of the underlying classical dynamic \cite{Rozenbaum2017}, the identification can be analytically shown for some particular systems. This is the case of uniformly hyperbolic linear maps on the tours \cite{OTOC_gato_PRL}, and the main steps of this key derivation are sketched below. Moreover, there is numerical evidence that the correspondence between $\Lambda$ and $\lambda$ holds in the chaotic regime for other physical systems like the Dicke model \cite{chavez2019quantum,sinha2021fingerprint} or a pair of coupled harmonic oscillators \cite{hashimoto2020cho}. For the bakers map, in Ref.~\onlinecite{lakshminarayan2019out} an analytical expression for the OTOC is derived, and  a factor 2 discrepancy with the expected Lyapunov growth is found, which the authors attribute to the non-linear peculiarities of the system.

For a system of $N$ coupled oscillators, the calculated OTOC growth rate in the large $N$-limit has been shown to agree with the numerically obtained classical Lyapunov exponent \cite{kolganov2022}. In a spin-chain with large spins, where a classical limit exists, the correspondence between the Lyapunov behavior and the OTOC growth rate has also been put in evidence \cite{craps2020lyapunov}.

Notwithstanding, there are many known exceptions that yield the behavior of Eq.~\ref{eq:qcOTOC} non-universal. There are systems considered chaotic by conventional criteria (i.e. non-integrable with random-matrix-like spectral statistics), which do not exhibit an exponential growth of the OTOC, but rather a power-law for $t < t^{*}$. Such is the case of systems without a clear classical analogue, like the spin-$1/2$ chains \cite{kukuljan2017,Motrunich2018,Riddel2019,FortesPRE2019,Shukla2022} (discussed  in Sect.~\ref{sec:spinchains}), as well as the Luttinger liquid \cite{moessner2017}. Conversely, an early-time exponential OTOC growth can be observed in polynomial billiards, which are examples of non-chaotic systems \cite{RozenbaumPRL2020}.

In addition, there are examples, like the inverted harmonic oscillator \cite{ali2020chaos,hashimoto2020exponential,morita2021extracting}, where the operator that evolves in time can be initially localized on an unstable fixed point in phase-space, resulting in an exponential growth of the OTOC for short times. Nevertheless, since the dynamics in these cases is not chaotic, the OTOC does not saturate, and therefore there is no scrambling but an oscillatory behavior typical of non-chaotic systems is observed for long times \cite{HummelPRL2019,XuPRL2020,kidd2021saddle}. Examples of these systems are addressed in Sect.~\ref{sec:saddle}.

Quantum maps on the torus are the simplest physical systems where all the essential features of quantum chaos are manifest. Therefore, they provide ideal numerical and analytical test-beds toward the goal of understanding the general properties of complex systems and the search for possible universal signatures in the three time-regimes.

The perturbed Arnol'd cat map relates the phase-space variables $(q,p)$ and $(q',p')$ of successive iterations through 
\begin{equation}
\label{clcatmap}
    \begin{array}{rcr}
p^{\prime}&=&p+q-2 \pi K \sin [2 \pi q] \,  \\
& &\\
q^{\prime}&=&q+p^{\prime}+2 \pi K \sin \left[2 \pi p^{\prime}\right] 
\end{array}
\quad \bmod 1 \, .
\end{equation}
This one-dimensional map has been shown to be particularly useful for quantum chaos studies \cite{Hannay1980,esposti2003mathematical} and we here use it in order to illustrate the features of the OTOC. The parameter $K$ represents the strength of the perturbation. This perturbation can be understood as a "kick" acting at each iteration of the map breaking the linearity of the unperturbed map. For small values of $K$, the Lyapunov exponent remains close to that of the unperturbed map, i.e. $\lambda\approx \ln [(3+\sqrt{5})/2]$.

The quantized version of the map is given by the unitary operator corresponding to the canonical (area preserving) transformation associated to the map. The periodic geometry of phase space results, after quantization, in a discrete Hilbert space with dimension $N$, and an associated effective Planck constant $\hbar_{\rm eff}=1/(2\pi N)$. The unitary operator that corresponds to the cat map ($M$) can be represented as the $N\times N$ matrix
\begin{equation}
\label{Upcat}
    \hU_{M}=e^{-\im 2 \pi\left[p^{2} / 2 N-K N \cos (2 \pi p / N)\right]} e^{-i 2 \pi\left[q^{2} / 2 N+K N \cos (2 \pi q / N)\right]} \, .
\end{equation}
The discrete variables $q$ and $p$ belong to the set $\mathbb{Z}_{N} =0,1,\ldots , N-1$ and are related  by a discrete Fourier transform. Thus, the map $\hat{U}_M$ can be efficiently implemented in numerical calculations. The structure of $\hU_M$, is closely related to that of Floquet systems obtained from periodically driven  Hamiltonians. This is also the case of the Harper \cite{ArtusoScholar} and the standard\cite{DimaScholar} maps. The link to Floquet systems has been recently highlighted in many domains such as the appearance of time crystals\cite{Else2016,Khemani2016,yao2017discrete}, topological insulators\cite{chiralfloquet,Anomalousfloquet}, as well as models of many-body localization \cite{PONTE2015196,Lazarides2015,PontePRL2015,abanin2016theory}.

For quantum maps, as well as for generic  Floquet systems there is no notion of energy, and therefore the thermal average looses meaning. However, the definition \eqref{otocdef1} for the OTOC becomes meaningful by trading the thermal average by the choice of the maximally mixed state $\rho_\infty\equiv \hat{I}_N/N$, with $\hat{I}_N$ the identity operator in a Hilbert space of dimension $N$. This is equivalent to  taking an infinite temperature limit in Eq.~\eqref{otocdef1}, thus working with the microcanonical  ensemble. As discussed in Sec.~\ref{subsec:definition}, the choice of infinite temperature and Hermitian operators results in ${\cal D}_{\hV\hW}(t)= {\cal I}_{\hV\hW}(t)={\rm Tr}\left\{\hW_t^2 \hV^2\right\}$. 

In order to compute the OTOC,  we  select the position and momentum operators, i.e. $\hat{W} \equiv \hat{Q}$ and $\hat V \equiv \hat{P}$. However, since position and momentum are not well-defined operators in the above specified Hilbert space,  an alternative definition in terms of the Schwinger unitary shift operators \cite{schwinger} can be used
\begin{equation}
\label{SCHW}
    \hat{\mathcal V}=\sum_{q \in \mathbb{Z}_{N}}|q+1\rangle \langle q| \, , \quad \hat{\mathcal U}=\sum_{q \in \mathbb{Z}_{N}}| q\rangle\langle q| \tau^{2 q} \, ,
\end{equation}
where $\tau=e^{\im \pi/N}$. From the generators of Eq.~\eqref{SCHW}  the following Hermitian operators can be defined
\begin{equation}
\label{defXP}
    \hat{Q}=\frac{\hat{\mathcal U}-\hat{\mathcal U}^{\dagger}}{2 \im}, \quad \hat{P}=\frac{\hat{\mathcal V}-\hat{\mathcal V}^{\dagger}}{2 \im} \ .
\end{equation}
In the semiclassical limit, $\hQ$ and $\hP$ fulfill the standard commutation relation of position and momentum operators. 

The shift operators are special cases of the Weyl translation operators $\hat{T}_{\vxi}=\hat{\mathcal V}^{\xi_{q}} \hat{\mathcal U}^{\xi_{p}} \tau^{\xi_{q} \xi_{p}}$, where $\vxi=(\xi_q,\xi_p)\in \mathbb{Z}^2$.
They have the following commutation properties
\begin{equation}
\label{transprop}
    \hat{T}_{\vxi} \hat{T}_{\vchi}=\tau^{\langle\vxi, \vchi\rangle} \hat{T}_{\vxi+\vchi},\left[\hat{T}_{\vxi}, \hat{T}_{\vchi}\right]=2 \im \sin \left(\frac{\pi}{N}\langle\vxi, \vchi\rangle\right) \hat{T}_{\vxi+\vchi} \, .
\end{equation}
where $\langle \ ,\ \rangle$ is the symplectic product.
More importantly, under quantum versions of a linear symplectic transformation, like cat maps, they transform classically, in the following sense 
\begin{equation}
\label{transM}
    \hat{T}_{M^{t} \xi}=\hat{U}_{M}^{\dagger t} \hat{T}_{\xi} \hat{U}_{M}^{t} \, ,
\end{equation}
where 
\begin{equation}
\label{matM}
M=\left(\begin{array}{cc} a&b\\ c&d \end{array}\right)\  \quad \text{and}\ \quad
M_t\equiv \left(\begin{array}{cc} a_t&b_t\\ c_t&d_t \end{array}\right)\ ,
\end{equation}
with ${\rm det}[M]={\rm det}[M_t]=1$.
Using  the $t$-times  evolved classical map $M_t$, Eq.~ \eqref{matM}, and the 
properties (\ref{transprop}) and (\ref{transM}) for the translation operators, with $\hat{Q}$ and $\hat{P}$ defined in (\ref{defXP}), an expression for the OTOC can be obtained\cite{OTOC_gato_PRL}  as  
\begin{equation}
\label{otoccatanalyt}
    {\cal C}_{\hP \hQ}(t)=\sin^2\left(\frac{\pi a_t}{N}\right) \, .
\end{equation}
For small times such that $a_t < N$ we have $a_t=e^{\lambda t}$ and thus

\begin{equation}
{\cal C}_{\hP \hQ}(t) \approx \frac{\pi^{2}}{N^{2}} \ e^{2 \lambda t} \, .
\end{equation}
Similarly, it can be shown that ${\cal F}_{\hP\hQ}(t)=\cos(2\pi a_t/N)/4$ and ${\cal D}_{\hP\hQ}(t) = {\cal I}_{\hP\hQ}(t)=1/4$. 

The derivation sketched above is important as it provides the main steps of one of the few analytical proofs that the exponential growth of the OTOC is given by the corresponding classical Lyapunov exponent $\lambda$ in a particular system. We stress that the derivation is restricted to linear automorphisms of the torus, in particular  the unperturbed version  ($K=0$) of the cat map, Eq. \eqref{clcatmap} . Numerical calculations at small $K$ indicate that the above relationship carries on, consistently with the insignificant change of $\lambda$ for small values of the perturbation strength. We remark that 
the discussion concerning the appropriate averaging of the Lyapunov exponent \cite{Rozenbaum2017}
is in fact avoided due to 
the  uniformly hyperbolic character of the cat map.
Moreover, the concepts of energy and temperature are rendered meaningless by the use quantum maps, 
leaving aside interesting features of the OTOC like the temperature-dependent bound of its growth-rate.  

%%%%%%%%%%%%%%%%%%%%%
\begin{figure}
    \centering
    \includegraphics[width=0.95\linewidth]{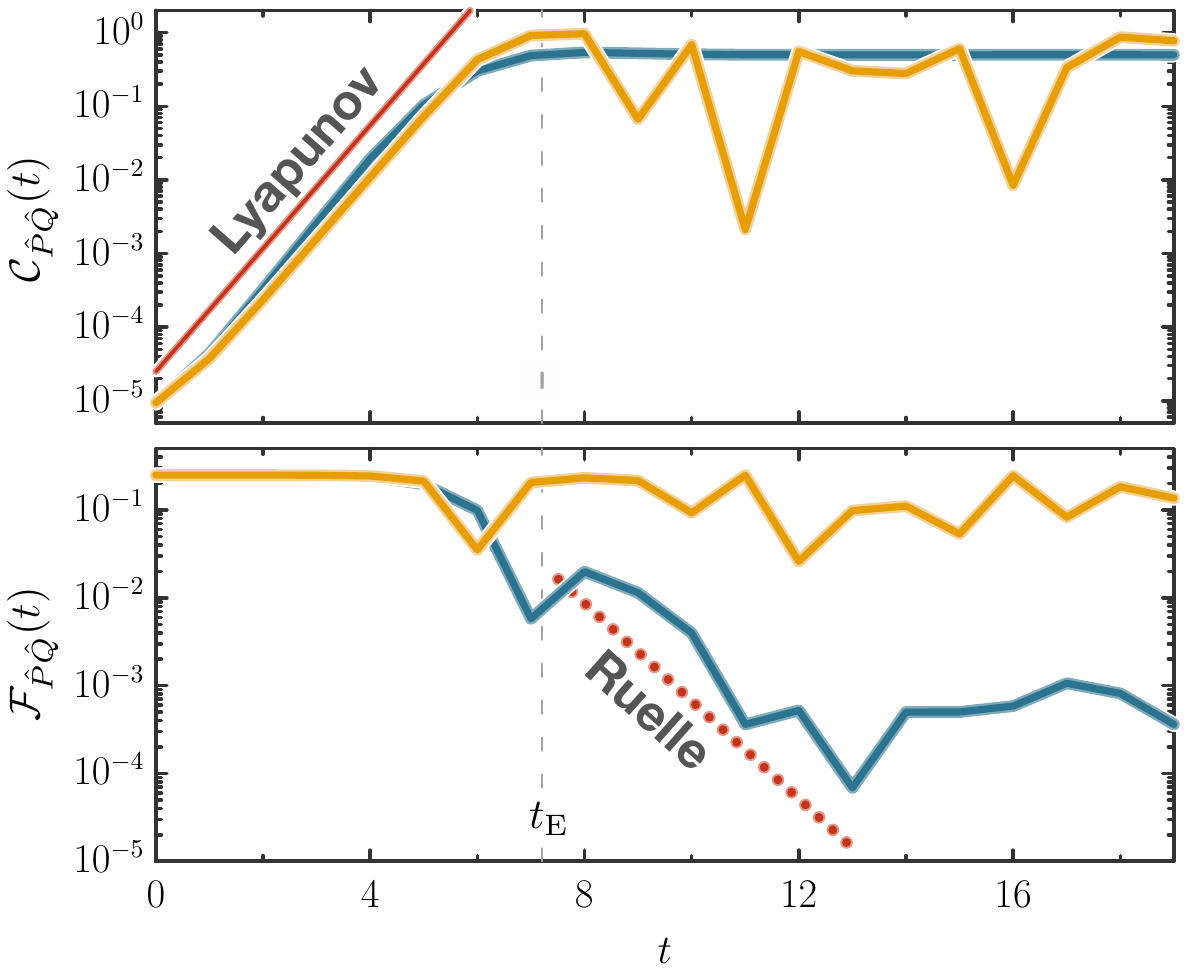} %%{figs/fig_otoccat_2panels.pdf}
    \caption{OTOC ${\cal C}_{\hP \hQ}(t)$ (top) and 4-point function ${\cal F}_{\hP \hQ}(t)$ (bottom) for the cat map with $N=1024$. The orange lines correspond to the analytic result for the unperturbed case ($K=0$), while the blue ones show the numerically obtained results for the perturbed case ($K=0.02$). The slope of the red curve in the top panel is given by twice the Lyapunov exponent $\lambda$ (obtained from the classical map). The dotted line in the bottom panel is proportional to $|\alpha_1|^{2t}$ where $\alpha_1$ is the Ruelle-Pollicott resonance with largest modulus, for the corresponding perturbed cat map. (Adapted from Ref.~\cite{OTOC_gato_PRL}, copyright 2018, American Physical Society.) \label{fig:otoccat1}}
\end{figure}

%%%%%
In Fig.~\ref{fig:otoccat1} (top) the analytical calculation of ${\cal C}_{\hP\hQ}(t)$ using Eq.~(\ref{otoccatanalyt}) (orange line) and a numerical  example of the perturbed case $K=0.02$  (dark blue line) are represented (both for $N=1024$), exhibiting the qualitative time behavior sketched in Fig.~\ref{Figtimescheme}. The Lyapunov regime is clearly observed for short times, up to the Ehrenfest time $\tE=\lambda^{-1}\ln{N} $ (marked with a dashed gray vertical line). 
Although we leave the discussion of longer times for the next sections, we would like to point out here some particular features of the cat map.
There is a clear difference between the unperturbed and perturbed cases in the long-time behavior. The unperturbed case does not saturate because, since cat maps transform translations into translations, the operators are never completely scrambled (this is also nicely shown in \cite{chen2018}). On  the other hand, when the nonlinear term is added the saturation to a constant value becomes evident. The fact that the unperturbed cat map does not scramble can also be observed in the behavior of ${\cal F}_{\hP \hQ}$, which remains close to its initial value during the Lyapunov regime and then, instead of decaying, it keeps on oscillating near that value, as shown in Fig.~\ref{fig:otoccat1} (bottom). The decay of  ${\cal F}_{\hP\hQ}$ for the perturbed case occurs after the Ehrenfest time and will be discussed in Sect.~\ref{sec:inttime}. 

In Refs.~\cite{OTOC_gato_PRL,FortesPRE2019} numerical evidence of short-time exponential growth of the OTOC with twice the classical Lyapunov exponent $\lambda$ was also numerically found for two other paradigmatic maps operating in the chaotic regime, namely the standard and the kicked Harper maps (see Fig.~\ref{harperlongtime}.b). 

\begin{figure}
    \centering
    \includegraphics[width=0.95\linewidth]{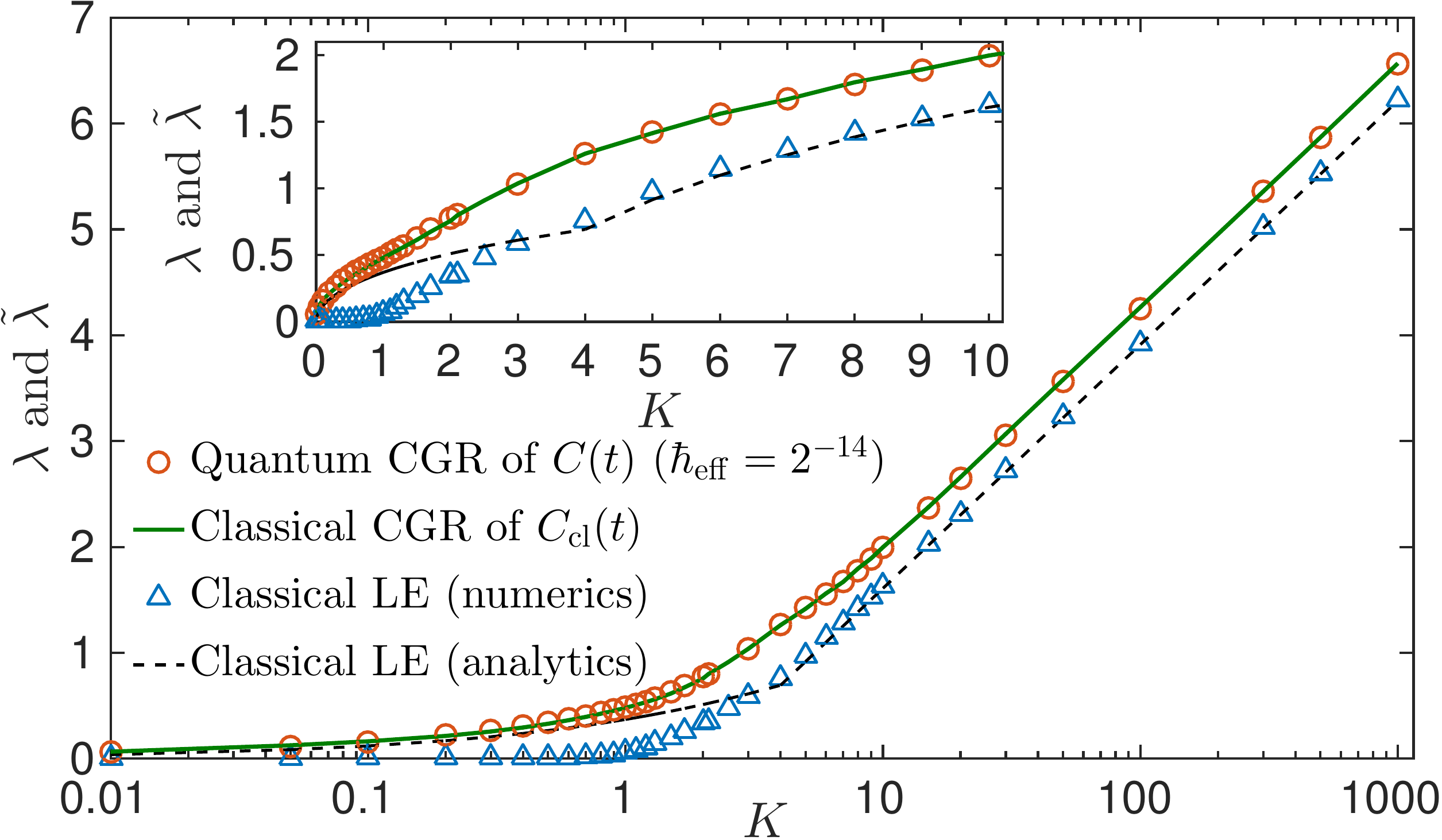} %%{figs/fig2Galitski.pdf}
    \caption{Four quantities are displayed in this figure in log-lin (main panel) and lin-lin (inset) scales, using for the kicked rotor map as a testbed. The triangles and black dashed lines represent the classical Lyapunov exponent obtained numerically and using a formula proposed by Chirikov in \cite{Chirikov1979}, respectively. The green solid line represents what the authors call commutator growth rate (CGR), obtained from the exponential growth of $C_{\mathrm{cl}}(t)=\hbar_{\text {eff }}^{2}\llangle\left(\Delta p(t)/\Delta x(0)\right)^{2}\rrangle$, where $\llangle \ldots \rrangle$ denotes a classical phase-space average. The quantum CGR is obtained by fitting an exponential law to the initial growth (before the Ehernefest time) of the numerically computed ${\cal C}_{\hV\hW}$.
    (Reproduced from Ref.~\cite{Rozenbaum2017}, copyright 2017, American Physical Society.) }
    \label{Figgalitski}
\end{figure}

The kicked rotor system offers another useful setup for studying the OTOC from a quantum chaos perspective, since varying the kicking strength allows to control the amount of chaos in phase space. In Ref.~\cite{Rozenbaum2017} the OTOC behavior is characterized, for times smaller than $\tE$, through a correlator growth rate (CGR) $\tilde{\lambda}= \Lambda/2$, and compared with the numerically determined classically Lyapunov exponent $\lambda$. For large values of the kicking strength $K$, it was found that $\tilde{\lambda}$ and $\lambda$ differ by an almost constant value of $\ln\sqrt{2}$ (see Fig.~\ref{Figgalitski}). The difference between the two exponents arises from the different order of taking averages and logarithms when computing the CGR and the Lyapunov exponent, but a clear relation between the two quantities, one classical and the other quantum, is soundly established. 

%%%%%%%%%%%%%%%%%%%%%%%%%%%%%
%%%%%%%%%%%%%%%%%%%%%%%%%%%%%%%%%%%%

Applying semiclassical theory to low dimensional billiard systems has been shown to lead to an exponential growth of the OTOC signed by the Lyapunov exponent\cite{JGMW2018}. This behavior comes after a much faster growth in the extremely short time regime. The time-window for the observation of the Lyapunov regime was found to be temperature dependent (larger for lower temperatures). These findings were numerically tested for the stadium billiard and are discussed in detail in Sect.~\ref{lowdim}. 

%%%
\subsection{Intermediate times}
\label{sec:inttime}
%%%%%%%%%%
After the initial growth, around the scrambling time $t^*$, the OTOC settles to an approximately constant value, with small oscillations in some cases. This not-too-well-defined interval, referred to as ``intermediate times'', is of special interest for strongly chaotic systems, because it can be  understood as the time between $t^*$ and complete relaxation. The scrambling time, where the information about an initial state is spread among the accessible space, can be defined by following the evolution of a complex system prepared in a pure state. The system will be considered to be ``scrambled'' if any subsystem has maximal entanglement entropy \cite{sekino2008fast}. 

In the case of a bounded one-particle system with classical analogue exhibiting chaotic dynamics, $t^*$ translates into the Ehrenfest time $\tE$. The latter is defined as the time that takes to a narrow coherent wave-packet to spread trough almost all available phase-space, and is thus given by $\tE = \lambda^{-1}\ln(a/\hbar)$, where the constant $a$ is set by the system-size and the initial wave-packet extension. (In the expression of $\tE$ used in the previous discussion of quantum maps $a=1/2\pi$). The Ehrenfest time establishes the upper limit for the applicability of Bohr correspondence principle, where the quantum evolution closely follows the corresponding classical distribution \cite{DimaScholar_Ehrenfest}. 

 A Wigner expansion of operators, in $\hbar$ for phase-space systems or in $1/S$ for spin systems with a large spins, allows to show that, with an underlying chaotic dynamics, the OTOC follows the classical spreading until the scrambling time \cite{cotler2018}. After $t^*$, quantum corrections become dominant, leading to the saturation of the OTOC. 

Chaos in a compact phase-space implies stretching and mixing. The stretching is related with the exponential separation of trajectories, whereas the mixing is attained when the stretching trajectories fold back unto themselves. While the stretching-dominated regime dictates the behavior of the dynamics in the short-time interval, mixing becomes relevant for intermediate times. While stretching is quantified by the Lyapunov exponent, mixing is quantified by the decay of correlation functions. 

For strongly chaotic systems, the decay of the correlation functions is exponential, with a rate given the Ruelle-Pollicot resonances (RPRs) \cite{pollicott1985rate,ruelle1986,ruelle1987resonances}. The RPRs are the isolated eigenvalues of the Koopman operator acting in a Banach space of generalized functions defined on the phase-space of the map \cite{Blank2002,Nonnenmacher2003}. When the functional space is restricted to the standard ${\cal L}^2$ of square-integrable functions, the Koopman operator is unitary. But outside this particular case, the Koopman operator spectrum consists of the unit eigenvalue, corresponding to the invariant distribution, a point spectrum, identified as the RPRs, plus a possible essential spectrum. In strongly chaotic systems, there is a finite gap between the largest (in modulus) RPR and the unit eigenvalue, which determines how correlations decay, governing the relaxation and the approach to equilibrium. 

\begin{figure}
    \centering
    \includegraphics[width=0.95\linewidth]{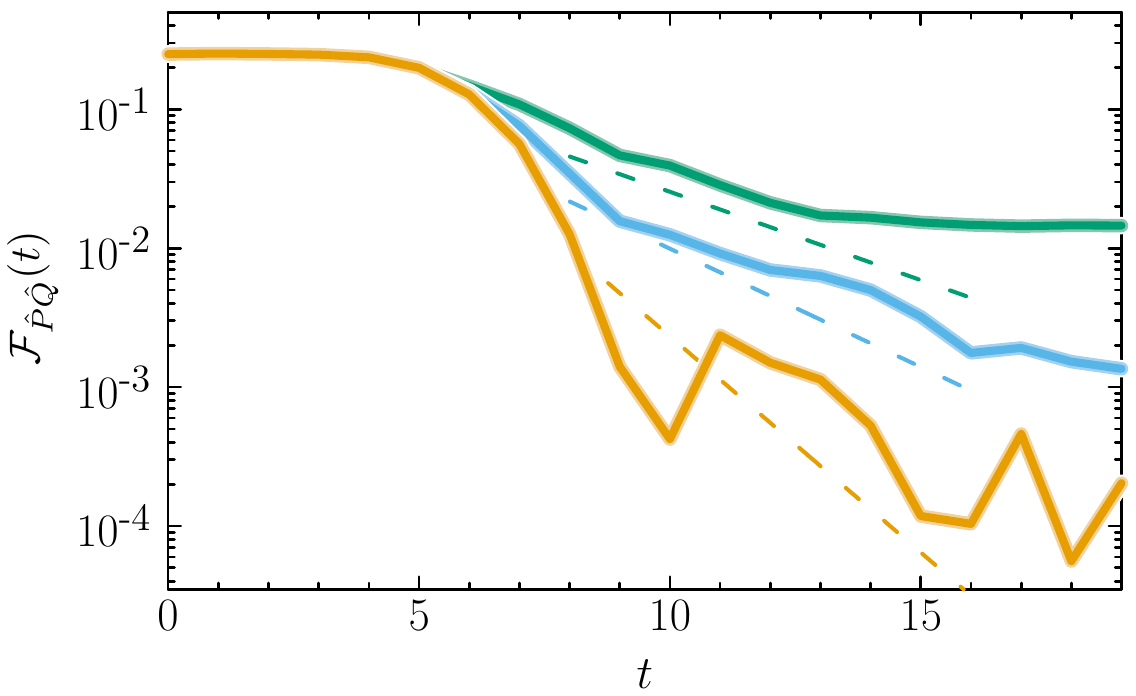} %%{figs/fig1_nacho_inset.pdf}  %%{fig1_alt_light.pdf}
    %{fig1-2.png}
    \caption{$|{\cal F}_{\hP\hQ}(t)|$ for the perturbed cat map for three values of perturbation strength $K=0.25$ (green), 0.275 (light blue), and 0.325 (orange). After the scrambling time the decay is approximately exponential. The dashed lines are proportional to $|\alpha_1|^{2t}$ where $\alpha_1$ is the largest-in-modulus Ruelle resonance for each value of the perturbation strength ($|\alpha_1| = 0.698$, 0.822, and 0.864, respectively). (Adapted from Ref.~\cite{OTOC_gato_PRL}, copyright 2018, American Physical Society.)}
    \label{figruellecat}
\end{figure}

For some kinds of strongly classically chaotic systems, a  correspondence between the spectrum of the (Gaussian coarse-grained) propagator of the density matrix and the RPRs can be established, upon appropriately taking the semiclassical and zero coarse-graining limits \cite{Nonnenmacher2003}. 
Then, by using an iterative method \cite{Blum2000,GarciaMata2004}, the largest RPRs can be efficiently obtained from the quantum propagator (e. g. the ones computed for Figs.~\ref{fig:otoccat1} and \ref{figruellecat}).
The influence of RPRs in chaotic systems can also be found in other related quantities like the purity (when considering bipartite systems, or open systems), the Loschmidt echo, and long-time behavior of spin echoes \cite{fine2005long}. Indeed, after the Ehrenfest time, these quantities saturate to a constant value, typically  dependent on the Hilbert space dimension. The way they approach to equilibrium is exponential and depends on  the RPRs  \cite{GarciaMata2003,GarciaMata2004,garciama2005}. 

In a study of chaos in black hole physics, Polchinski \cite{polchinski2015} signaled two time-regimes for the OTOC evolution, characterized by two exponential laws: the early exponential growth ("Lyapunov") for short times, and the later exponential decay governing the approach to equilibrium ("Ruelle"). Fig. 2 of Ref.~\onlinecite{polchinski2015} can be related to  Fig.~\ref{Figtimescheme} above (up to the long-time oscillations) by identifying "Lyapunov" with "short-time" and "Ruelle" with intermediate time. 
 
In the case of the OTOC for quantum maps, where the approach of ${\cal C}_{\hP\hQ}(t)$ to saturation can be read from the decay of ${\cal F}_{\hP\hQ}(t)$, it was found \cite{OTOC_gato_PRL} that ${\cal F}_{\hP\hQ}$ falls exponentially  as $|\alpha_1|^{2t}$, where $\alpha_1$ is the RPR with largest modulus (verifying $|\alpha_1|<1$). These results are illustrated for the perturbed cat map in Fig.~\ref{fig:otoccat1} (bottom), where it can be observed that ${\cal F}_{\hP\hQ}$ is approximately constant up to the scrambling (Ehrenfest) time and then it decays exponentially. The perturbed cat map is particularly well suited to observe this effect since $|\alpha_1|$ depends strongly  on the perturbation strength $K$, allowing to unequivocally demonstrate the role of the RPRs for the time-dependence of ${\cal F}_{\hP\hQ}$. The decay of ${\cal F}_{\hP\hQ}$ is shown in Fig.~\ref{figruellecat} for three different values of perturbation strength for the cat map \eqref{Upcat}, and the agreement with the behavior dictated by the corresponding $|\alpha_1|^{2t}$ is clearly visible. 

We conclude this chapter by stressing that the two main ingredients in the dynamics of a strongly chaotic classical system, which are the exponential separation of trajectories (butterfly effect encoded in the Lyapunov exponent) and the mixing (manifested in the exponential decay of correlation functions), present analogous effects in the behavior of the OTOC of the corresponding quantum system, for small and intermediate times, respectively.

\subsection{Long times}
As stated in the previous chapters, the initial interest on the OTOC was completely focused on the possibility to relate its initial exponential growth to chaos. Later understanding that the previous relationship is not straightforwardly universal motivated further studies and the consideration of different time-regimes. 

On the one hand, in the case of strongly chaotic systems, the relaxation after the scrambling time leads to a constant value of the OTOC, and several works addressed the dependence on system parameters of the long-time saturation values of ${\cal C}_{\hW\hV}(t)$ and ${\cal F}_{\hW\hV}(t)$ \cite{hashimoto2017,JGMW2018,rammensee2018,huang2019finite,sinha2021fingerprint,Markovic2022}. On the other hand, it was realized that the long-time behavior of the OTOC strongly depends on the dynamics (chaotic or regular), and that a deep insight can be obtained from the study of the fluctuations in the long-time regime. Indeed, very precise measures of quantum chaos can be defined from the long-time behavior of the OTOC, and the analysis of its oscillations allows to quantify the transition from regular to chaotic dynamics in a accordance with other traditional chaos indicators \cite{FortesPRE2019}. We discuss in detail these measures in Sec.~\ref{sec:QCIFLTB}, and we summarize below some salient observations for the long time-regime.

In one-body systems that have a completely chaotic classical dynamics, the saturation value of the of the OTOC scales linearly with the temperature and the system-size for the case where where $\hW=\hat{X}$ and $\hV=\hat{P}_X$ (see Eq.~\eqref{eq:CLongT}), but different temperature scaling could be obtained for other choices of operators. In Ref.~\onlinecite{Markovic2022} it is described how the temperature scaling of the OTOC saturation value might change according to the integrability of the one-body dynamics and the complexity of the chosen operators.  

Studies of the Dicke model  \cite{sinha2021fingerprint} and the driven Bose-Hubbard dimer have shown that the the exponential growth driven, either by genuine chaos or by a saddle in phase-space, can be efficiently distinguished by the long time-behavior of the OTOC \cite{kidd2021saddle}. While chaotic dynamics leads to saturation, the case where the behavior is dominated by the presence of a saddle leads to long-time oscillations of the OTOC.  

The long-time OTOC behavior of chaotic many-body systems leads to a size-dependent saturation, which can be analyzed semi-classically in terms of many-body post-Ehrenfest quantum interferences \cite{rammensee2018}. Such a behavior amounts for an indefinite OTOC growth in the thermodynamic limit. Numerical and analytical results for the kicked Ising model confirm this expectation in the chaotic regime, but exhibit a genuine plateau (outlasting the thermodynamic limit) for the integrable case and special choices of the operators \cite{kukuljan2017}. In many-body localized systems, where the scrambling is considerably suppressed, the long-time OTOC saturation is system size independent \cite{He2017}.

In Ref.~\onlinecite{WangESQPT2019} the long-time behavior of the OTOC was proposed as a possible order parameter for the excited state quantum-phase-transition taking place in the Lipkin-Meshkov-Glick model after a sudden quench.

The long-time behavior of the OTOC was addressed in Ref.~\onlinecite{huang2019finite} by studying how the saturation of ${\cal F}_{\hW\hV}(t)$ scales with the system-size for energy-preserving time-independent chaotic Hamiltonians. In particular, it was found (semi analytically) that for local operators the scaling is an inverse polynomial in the system size. 

When considering operators acting on separate subsystems, one can define a bipartite OTOC, which has interesting properties. In particular the long-time behavior of this bipartite OTOC can be related to operator entanglement properties \cite{styliaris2021information}.

Ref.~\onlinecite{wang2021microscope} considered a generalized OTOC, defined over points in phase-space. When the OTOC is computed on a grid of Gaussian states and its long-time value is plotted as a function of the position in phase space, it is obtained a density-plot image allowing to differentiate regular and chaotic regions with a resolution given by the effective Planck constant.  

%%%%%%%%%%%%%%%%%%%%%%%%%%%%%%%%%%%%%%%%%%%%%%%%%%%%%%%%%%%%%%
\begin{figure}
    \centering
    \includegraphics[width=0.95\linewidth]{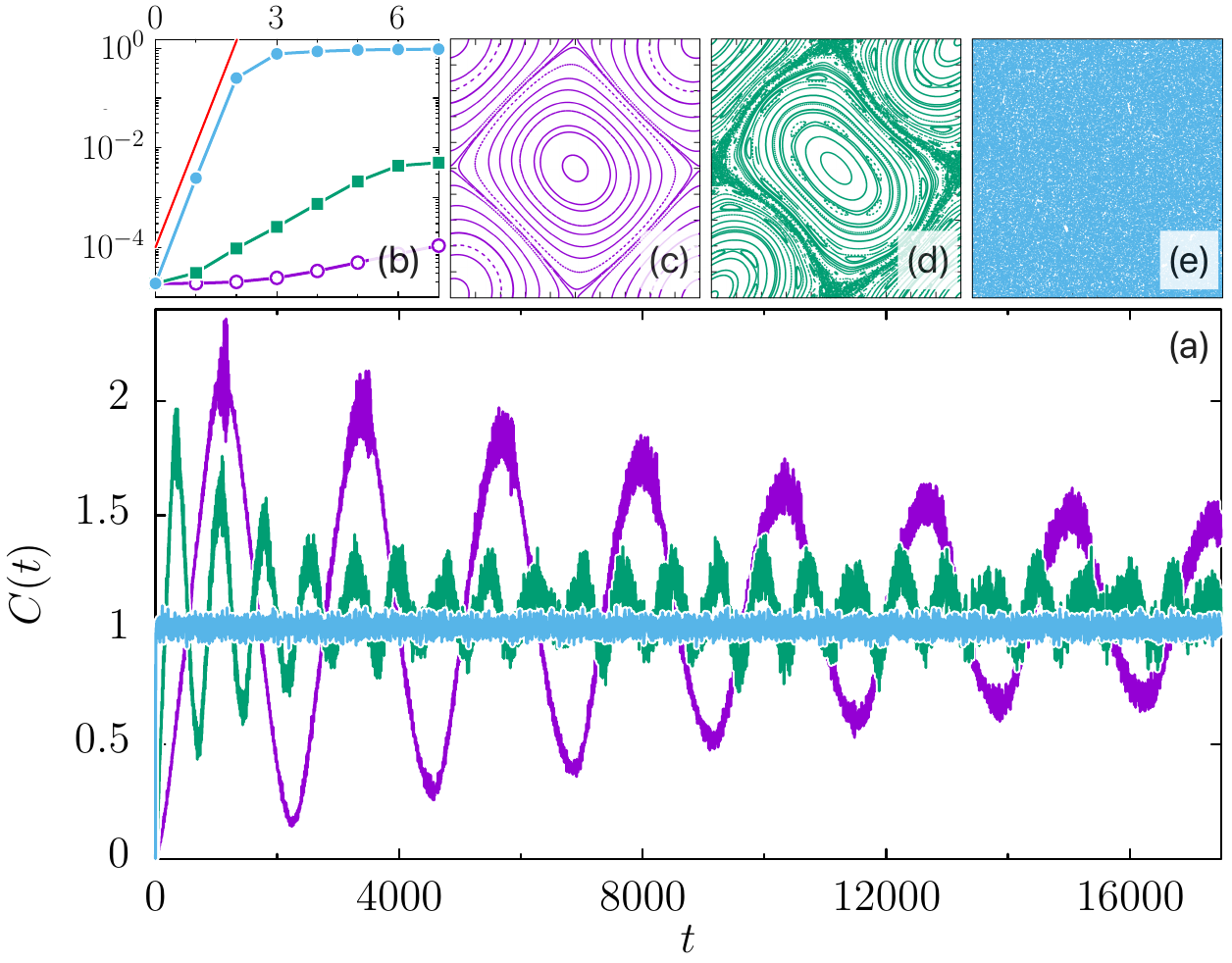}  %%{figs/OTOC_Harper_longtime.pdf}  %% 
    \caption{Main figure (a): Time-dependence of the OTOC for the Harper map, for different values of the kicking parameter $K$, over a large time-interval. Panel (b): blow-up of the previous figure, in a logarithmic scale, showing the initial growth of the OTOC. The red line has a slope given by the Lyapunov exponent $\lambda$, obtained from the classical dynamics of the map. Panels (c),(d) and (e) show the phase-space portraits for the three curves in (a), with the corresponding color. The least chaotic case (smallest $K$, panel (c)) is characterized the the largest oscillations, exhibiting one dominating frequency. In the most chaotic case (largest $K$, panel (e)) the OTOC oscillates rapidly with small amplitudes around the saturation value. (Reproduced from Ref.~\cite{FortesPRE2019}, copyright 2019, American Physical Society.)}
    \label{harperlongtime}
\end{figure}
%%%%%%%%%%%%%%%%%%%%%%%%%%%%%%%%%%%%%%%%%%%%%%%%%%%%%%%%%%%%%%

Leaving aside the issue of the saturation value of the OTOC, it has been found the fluctuations of the OTOC at large times, beyond the scrambling time can contain crucial information concerning the  amount of chaos in a system. In many cases where chaos can be tuned by one parameter, a generic behavior of the OTOC can be observed. As an example, in  Fig.~\ref{harperlongtime} it is shown the case of the kicked Harper map, where the kicking strength $K$ controls the transition from integrable to chaotic dynamics. Then for very small $K$ the system is regular (see the Poincar\'e surface of section in panel (c)), and for very large $K$ the system is completely chaotic (panel (e)).
In the main panel (a) the time-dependence of the OTOC corresponding to the kicking parameters of the panels ((c) to (e) with the corresponding colors) is shown, and in panel (b) a close-up of the data for the short times is presented. It can be observed that the regular regime is associated to very large amplitude oscillations, which are characterized by approximately only one frequency. The decrease of the amplitude of the oscillations with $K$, is evident from the picture. This behavior also characterizes other systems, like the quantum standard map and spin chains \cite{FortesPRE2019}, as well as a one-dimensional Bose gas \cite{HummelPRL2019}. Notably, such generic behavior can also be observed in very the short spin chains accessible in NMR experiments \cite{Li2017}. 
 
The significant difference of the fluctuating patterns of the OTOC, at long times, for regular and chaotic systems, has allowed to design measures or indicators of chaos that accurately reproduce the transition from regular to chaotic. We describe some of them in Sect.~\ref{sec:QCIFLTB}.
%%
%%
%%%%%%%%%%%%%%%%%%%%%%%%%%%%%%%%%%%%%%%%%%%%
\section{OTOC and quantum chaos} 
 The relationship \eqref{eq:qcOTOC}, valid for classically chaotic systems, provides a clear connection of the OTOC with quantum chaos, as the latter concerns the study of quantum systems with an underlying classically chaotic dynamics. The possibility of using the exponential growth of the OTOC as a measure of quantum chaoticity, especially in systems without classical limit and in many-body systems, provided a strong incentive for investigating the universality of the OTOC in different areas of Physics, ranging from High Energy to Condensed Matter, and Mesoscopic Physics.
 
In trying to establish the generality of the relationship  \eqref{eq:qcOTOC}, it is important to test it in the cases where the classical analogue exists. We thus consider in this section the one-body and the many-body semi-classical approaches to the OTOC, also addressing the case of systems without a classical analogue. We then present the chaos signatures in the long-time behavior of the OTOC and discuss the case of integrable systems exhibiting an instability in their classical dynamics.

\subsection{Low-dimensional classically chaotic systems}
\label{lowdim}
The paradigmatic case of a billiard, that is, one particle with its motion limited by a two-dimensional confining potential, has been used for OTOC studies in low-dimensional systems \cite{hashimoto2017,JGMW2018,rozenbaum2019}. Choosing $\hat X$ and $\hat P_X$ as operators, each of the terms of Eq.~\eqref{otocdef2} can be written as an energy and space integral
\begin{align}
{\cal O}(t) &= - \frac{1}{\pi Z} \int  \d \ve \ \d \rvp \ \d \rv \ e^{-\beta \ve} \ \nonumber \\
& \ \ \ \ {\rm Im} 
\left\{G(\rv,\rvp;\ve)\right\} \ O(\rvp,\rv;t) \, .
\label{eq:OTOCImGF}
\end{align}

\begin{itemize}

\item ${\cal O}(t) = {\cal D}_{\hat P_X\hat X}(t)$, \, ${\cal I}_{\hat P_X\hat X}(t)$, or ${\cal F}_{\hat P_X\hat X}(t)$, where the operator sub-index is herewith eliminated in order to lighten the notation.

\item $O(\rvp,\rv;t) = \langle \rvp \left| {\hat O}_t \right| \rv \rangle$ are matrix elements, in the basis of position eigenstates, of the operator ${\hat O}_t = \hP_X \hX^{2}_t \hP_X$,  $\hX_t\hP_X^2\hX_t$, or $\hX_t \hP_X \hX_t \hP_X$, respectively.

\item $Z={\cal A}m/(2\pi \hbar^2 \beta)$ is the partition function for a spinless particle of mass $m$ in a billiard of area ${\cal A}$.

\item ${\rm Im}\left\{G(\rv,\rvp;\ve)\right\}$ stands for the imaginary part of the Green function. 

\end{itemize}

The Green function $G(\rv,\rvp;\ve)$ is defined as the Fourier transform of the propagator, itself a matrix element of the evolution operator 
\begin{equation}
K(\rvp,\rv;t) = \langle \rvp \left| \hU_t \right| \rv \rangle \, . 
\label{eq:propagator}
\end{equation}

In the fully chaotic case the calculation of the different $O(\rvp,\rv;t)$ can be addressed using a semi-classical approach, by expressing the propagator as a sum over classical trajectories:
\begin{align}
K_{\rm sc}(\rvp,\rv;t) & = \left(  \frac{1}{2\pi \im {\hbar}}\right) \sum_{s(\rv,\rvp;t)}
C_{s}^{1/2} \nonumber \\
&  \ \ \ \ \exp{\left[\tfrac{\im}{\hbar}R_{s}(\rvp,\rv;t)-\im \tfrac{\pi}{2}\mu_{s}\right]  } \, . 
\label{eq:SCpropagator}
\end{align}
\begin{itemize}

\item $s(\rv,\rvp;t)$ label the isolated classical trajectories joining the points $\rv$ and $\rvp$ in a time $t$, upon which the sum is performed.

\item \mbox{$R_{s}(\rvp,\rv;t)=\int_{0}^{t} \d \tau \mathcal{L}$} is the Hamilton principal function, $\mathcal{L}$ is the Lagrangian of the system, the integration is performed along the classical path, and $\mu$ is the Maslov index counting the number of conjugate points. 

\item $C_{s}=\left|  \det\mathcal{B}_{s}\right|$, with 
$(\mathcal{B}_{s})_{ab}=-\partial^{2}R_{s}/\partial r_{a}^{\prime} \partial r_{b}$, the primed (un-primed) variables correspond to the final (initial) position, while $b$ and $a$ stand for the Cartesian coordinates.

\end{itemize}
The semi-classical expansion remains valid over times that greatly exceed the Ehrenfest time \cite{tomsovic1991}.

The semi-classical approximation to the matrix element $D(\rvp,\rv;t) $ is given as a double sum over classical trajectories by
\begin{align}
D_{\rm sc}(\rvp,\rv;t)  &  = - \frac{1}{\left(2\pi \hbar\right)^2}  \int \d \rv_{1}
\sum_{s_{2}(\rv_{1},\rvp;t)} \ \sum_{s_{1}(\rv,\rv_{1};t)} 
C_{s_{2}}^{1/2} \ C_{s_{1}}^{1/2} \nonumber \\
& \times
\left\{P_{X,s_{2}}^{\rm f} \left( X_{1} \right)^{2} P_{X,s_{1}}^{\rm i}\right\} \,
\exp\left[  \frac{\im}{\hbar}\left(R_{s_{1}}(\rv_{1},\rv;t)
\right. \right.
\nonumber \\ &
\left. \left.
-R_{s_{2}}(\rvp,\rv_{1};t)\right)  -\im \frac{\pi}{2}\left( \mu_{s_{1}}-\mu_{s_{2}}\right)  \right]   \, ,
\label{eq:O3semi}
\end{align}

\begin{itemize}

\item $X_1=\rv_1.\hat{\bf e}_X$ and $P_X=\pv.\hat{\bf e}_X$, with $\hat{\bf e}_X$ the unit vector in the $X$-direction.

\item The indices ${\rm i}$ and ${\rm f}$ refer, respectively, to the initial and final condition of the corresponding trajectory. 

\end{itemize}

\begin{figure}
    \centering
\centerline{\includegraphics[width=0.48\textwidth]{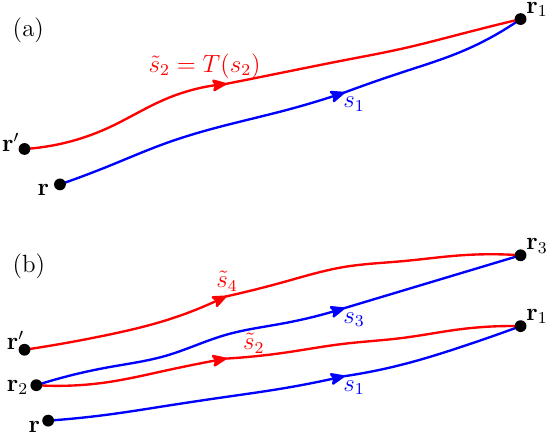}} %%{figs/complete.pdf}} %%{t3g}}
\caption{(a): Graphical representation of $D_{\rm sc}(\rvp,\rv;t)$ according to Eq.~\eqref{eq:O3semi} for the case in which the trajectories $s_1$ and $s_2$ remain close to each other. The label ${\tilde s}_{2}=T(s_{2})$ stands for the time reversed trajectory of $s_{2}$, and $\rv_1$ represents the intermediate integration position. The color blue (red) is used for trajectories whose Hamilton principal function appears with a plus (minus) sign in the phase term of Eq.~\eqref{eq:O3semi}. (b): Graphical representation for the semi-classical approximation of the other OTOC components $I_{\rm sc}(\rvp,\rv;t)$ and $F_{\rm sc}(\rvp,\rv;t)$, where four nearby trajectories are involved. The color convention in the same as in panel a, while $\rv_1$, $\rv_2$, and $\rv_3$ represent the intermediate integration positions. (Adapted from Ref.~\cite{JGMW2018}, copyright 2018, American Physical Society.)}
    \label{fig:trajectories}
\end{figure}

Independent trajectories where the corresponding phases are unrelated average out their contribution upon the spatial integrations. Therefore, the dominant contributions stem from terms associated with trajectory pairs in which the corresponding phases are related. The most obvious connection is when $\tilde{s}_{2} = T(s_2)$ (the time-reversal symmetric of the trajectory $s_{2}$) remains close to $s_{1}$. Fig.~\ref{fig:trajectories}.a provides a graphical representation of the dominant pairs of trajectories contributing to $D_{\rm sc}(\rvp,\rv;t)$ in Eq.~\eqref{eq:O3semi}. The semi-classical expressions for $I_{\rm sc}(\rvp,\rv;t)$ and $F_{\rm sc}(\rvp,\rv;t)$ (not presented, see Ref.~\cite{JGMW2018}) are more complicated than that of $D_{\rm sc}(\rvp,\rv;t)$, as they are given by sums over four classical trajectories. The generic form of the dominant terms is sketched in Fig.~\ref{fig:trajectories}.b. The classical trajectories drawn in Fig.~\ref{fig:trajectories} can generically be viewed as special OTOC contours (or modes) presenting particularly large correlations \cite{Ueda2018,stanford2022}.

While semi-classical approaches have been developed for the temperature-dependent stationary case \cite{ullmo1997,ullmo1998,ullmo2008}, the time and temperature dependence in Eq.~\eqref{eq:OTOCImGF} necessitates a more elaborate treatment, requiring a mixed representation involving the energy-dependent Green function and the time-dependent propagator. A considerable simplification can be obtained in the energy integration of Eq.~\eqref{eq:OTOCImGF}, by using the free-space Green function. This approximation yields that the dominant terms of the semi-classical expansion are for the case in which $\rv$ and $\rvp$ are close to each other (as assumed in the discussion of Eq.~\eqref{eq:O3semi}).

The classical limit (leading contribution of order $\hbar^0$) is the same for the three OTOC components ${\cal D}(t)$, ${\cal I}(t)$, and ${\cal F}(t)$. It can be obtained, through a strict diagonal approximation identifying the classical trajectories $s_1$ and $\tilde{s}_{2}$ in Fig.~\ref{fig:trajectories}.a for $D_{\rm sc}(\rvp,\rv;t)$  (together with the n of $s_3$ and $\tilde{s}_{4}$ in Fig.~\ref{fig:trajectories}.b for $I_{\rm sc}(\rvp,\rv;t)$ and $F_{\rm sc}(\rvp,\rv;t)$), as
\begin{equation}
{\cal O}_{\rm cl}(t) = \frac{\beta}{2\pi {\cal A} m} \int \d \rb  \ \d \pb 
\ \exp{\left[-\beta\frac{ \pb^2}{2 m} \right]} \ 
\{ {\bar P}_{X}^{2} \ X^{2}(\rb,\pb;t) \}  \, .
\label{eq:O3class}
\end{equation}

\begin{itemize}
    \item $X(\rb,\pb;t)$ represents the $X$-component of the particle position at time $t$ resulting from an initial condition defined by the point $(\rb,\pb)$ of the phase-space upon which the integration is carried out.
    \item $\beta/(2\pi {\cal A} m) = [(2\pi \hbar)^2 Z]^{-1}$, and thus the above expression simply represents the thermal average of ${\bar P}_{X}^{2} \ X^{2}(\rb,\pb;t)$.  
\end{itemize}

From \eqref{eq:O3class}, it follows that ${\cal C}_{\rm cl}(t)=0$, which is an obvious result, since the finite value of ${\cal C}(t)$ arises from the operators' non-commutativity, which is a purely quantum concept. 

The next-order corrections in $\hbar$ of the OTOC components are obtained by calculating $\delta {\cal O}(t) = {\cal O}(t) - {\cal O}_{\rm cl}(t)$. To order $\hbar^2$, we have $\delta {\cal D}(t)=0$ and $\delta {\cal I}(t)=\delta {\cal F}(t)$. Assuming a uniformly hyperbolic system, the exponential divergence of nearby trajectories characterized by a Lyapunov exponent $\lambda$ leads to the semi-classical approximation of the OTOC \cite{JGMW2018}
\begin{equation}
{\cal C}_{\rm sc}(t) = \frac{\beta^2 \hbar^2}{64 \pi m^2}  \int  \d \pb 
\ \exp{\left[-\beta\frac{ \pb^2}{2 m} \right]} \ 
\left\{ e^{2\lambda t} \ {\hat P}_{X}^{2} \right\}  \, .
\label{eq:Csemiclass}
\end{equation}

As expected, ${\cal C}(t)$ scales with $\hbar^2$. Since the Lyapunov exponent is energy-dependent (and thus $|\pb|$-dependent), the simple OTOC exponential growth of Eq.~\eqref{eq:qcOTOC} can only be achieved under special conditions. In particular, for a billiard, for sufficiently low temperatures and for not too long times, 
\begin{equation}
\frac{{\cal C}_{\rm LT}(t)}{\hbar^2} \varpropto 
\exp{\left[\sqrt{3} \ \lambda_{\rm g} {\tilde v} t \right]}  \, ,
\label{eq:CLowT}
\end{equation}

\begin{itemize}
    \item  $\lambda_{\rm g} = \lambda t/L$ is a purely geometrical Lyapunov exponent, and $L = (|\pb|/m)t$ is the trajectory length. 
    \item ${\tilde v} = \left\langle V_X^2 \right\rangle^{1/2} = (\kb T/m)^{1/2}$ is the root-mean-square for the $X$-component of the velocity for a free two-dimensional particle in contact with a thermostat at a temperature $T$. 
\end{itemize}

Leaving aside the hypothesis of a uniform hyperbolic dynamics (where the exponential growth has the same Lyapunov exponent within each energy-shell of phase-space) used to obtain Eq.~\eqref{eq:Csemiclass}
raises the question of the different ways of performing a phase-space average of the Lyapunov exponent. This important issue have been discussed for the case of the Loschmidt echo \cite{Silvestrov2003}, as well as for the OTOC \cite{Rozenbaum2017,rozenbaum2019}.

For long times, the OTOC saturates to a value
\begin{equation}
{\cal C}_{\rm s} \varpropto m a^2  \kb T \, ,
\label{eq:CLongT}
\end{equation}
that scales with the temperature and the area of the billiard, but is independent of $\hbar$ \cite{hashimoto2017}. 

%%%%%
\begin{figure}
    \centering
\centerline{\includegraphics[width=0.48\textwidth]{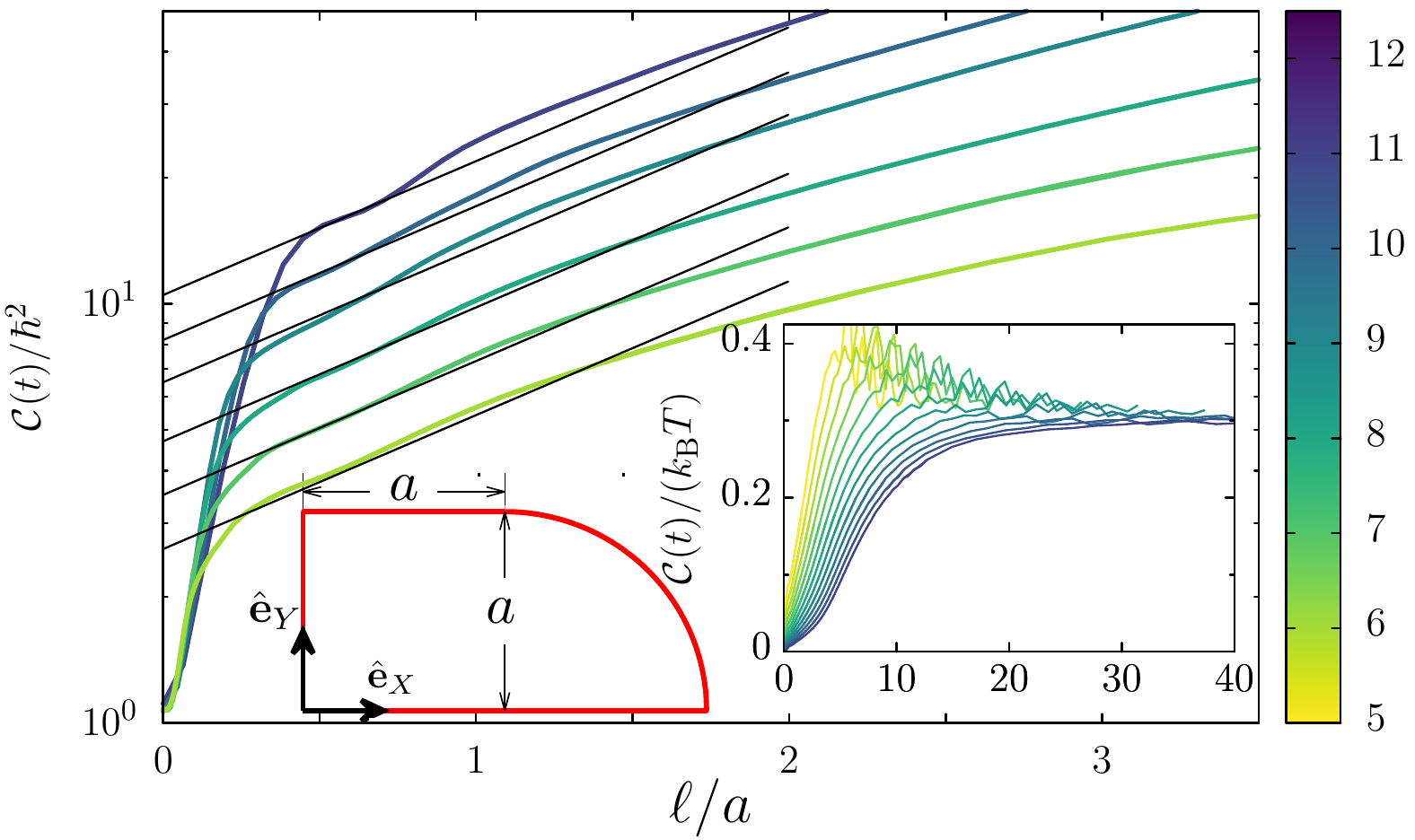}} %%{figs/stadium.pdf}} %%{t3g}}
\caption{Numerically obtained OTOC (in a logarithmic scale) as a function of the scaled time (length) $\ell = {\tilde v} t$ (in units of $a$), with $ {\tilde v} = (\beta m)^{-1/2}$ the mean-squared $X$-velocity component, for the desymmetrized stadium sketched at the bottom.  The color code indicates the temperature scale, expressing $\kb T/E_0$ in a $\log_2$ basis, with $E_0=\hbar^2/(ma^2)$. The black straight lines describe the corresponding exponential growth $e^{\sqrt{3} \lambda_{\rm g} \ell}$ applicable in an intermediate time-window. Inset: OTOC scaled with the temperature in a large $\ell/a$ interval showing the long-time saturation. (Adapted from Ref.~\cite{JGMW2018}, copyright 2018, American Physical Society.)}
    \label{fig:stadium}
\end{figure}
%%%%%%%%

Fig.~\ref{fig:stadium} presents the quantum numerical calculation of the OTOC for the desymmetrized stadium sketched in the inset, as a function of the scaled time (or length) $\ell = {\tilde v} t$ at various temperatures $T$ (indicated by the color scale). The three main time-regimes discussed in Sec.~\ref{sec:time_reg} are visible: short time, intermediate time, and long time. 

Within the short time regime, we distinguish an initial quadratic take-off (on $t$ or $\ell$) characteristic of quantum perturbation theory, followed by a rapid growth, turning into a $\ell$-window with an exponential increase of the OTOC. The previous semi-classical approach applies to this last interval, but not to the perturbative or rapid growth intervals, since they correspond to times much smaller than that of the first collision with the boundaries, and therefore the exponential divergence of classical trajectories is not effective. 

The exponential increase is well fitted (black solid lines) by Eq.~\eqref{eq:CLowT} when using $\lambda_g=0.425 a^{-1}$ (applicable to the chosen billiard). The $\ell$ (or time) window of the exponential increase enlarges upon lowering the temperatures and shrinks when raising the temperature, until disappearing. The thermal washout of the energy-dependent Lyapunov exponent might be at the origin of the non-identification of a window of exponential growth in other simulations of the stadium billiard \cite{hashimoto2017}. According to the semi-classical prediction \eqref{eq:CLowT}, the OTOC growth-rate is $\Lambda=\sqrt{3} \ \lambda_{\rm g} {\tilde v}$, and thus scales as $T^{1/2}$. Such a result is compatible with the bound on the OTOC growth-rate $\Lambda \le 4\pi \kb T/\hbar$ proposed in Ref.~\cite{maldacena2016}, outside the extreme low-temperature case in which $\kb T$ is of the order of the ground state of the billiard. 

The saturation of the OTOC for long times, expected for a finite-size system around the Ehrenfest time, is proportional to temperature (see inset), in agreement with Eq.~\eqref{eq:CLongT}. The semi-classical estimation of the proportionality constant is problematic, due to the different possible pairings \cite{gutierrez2009} and the effect of trajectory loops \cite{sieber2001,Gutkin2010}. The intermediate time-regime is characterized by oscillations of the OTOC as a function of $\ell$, reflecting the dynamics of the billiard and the signature of the periodic-orbit corrections \cite{JGMW2018}.

The semi-classical approach above presented assumes a completely chaotic classical dynamics. The case of an integrable  dynamics has been numerically investigated \cite{hashimoto2017}. For a circular billiard, the initial OTOC growth has not been found to follow a simple exponential increase, while the long-time behavior is characterized by oscillations, instead of a saturation.    
%%%%%%%%%%%%%%%%%%%%%%%%%%%%%%%%%%%%%%%%%%%%%%%%%%%%%
\subsection{Many-body semi-classics}%
%%%%%%%%%%%%%%%%%%%%%
Within the goal of characterizing the many-body quantum dynamics, several numerical calculations of the OTOC have been implemented in a variety of interacting systems 
(\cite{kukuljan2017,Motrunich2018,Riddel2019,FortesPRE2019,Swingle2019,Borgonovi2019,craps2020lyapunov,Shukla2022} among others, for a review see \cite{swingle2022}), but few analytical tools are at our disposal. Among them, the many-body semi-classics is particularly promising since for some classes of systems it can be developed in analogy with the standard one-particle case \cite{engl2014}.  

For a bosonic $N$-particle system in a lattice, the many-body version of the propagator can be expressed as in the one-body case, by Eq.~\eqref{eq:propagator}, where

\begin{itemize}
    \item $\hU_t$ is the evolution operator in Fock space,
    
    \item $| \rv \rangle$ and $| \rvp \rangle$ represent, respectively, initial and final localized coherent states (or more generally, quadratures).
\end{itemize}

The semi-classical approximation to the many-body propagator has an analogous form to its one-body counterpart \eqref{eq:SCpropagator}, with changes in the meaning of the various factors. 

\begin{itemize}
    \item The sum runs over all time-dependent solutions $s$ of the classical (mean-field) equations of motion (governed by the classical limit  of the Hamiltonian) with the boundary conditions given by $| \rv \rangle$ and $| \rvp \rangle$.
    
    \item $R_{s}(\rvp,\rv;t)$ is the Hamilton principal function along classical path $s$, and the weights $C_{s}$ reflect the corresponding classical stability.
    
    \item The small parameter who plays the role of Planck constant is $\hbar_{\rm eff}=1/N$. 
    
\end{itemize}

The mean-field dynamics of the $N$-particle system is difficult to characterize. But the assumption of an uniformly hyperbolic, chaotic
dynamics, where the exponential separation of trajectories has the same Lyapunov exponent $\lambda$ at any phase-space point \cite{rammensee2018}, brings us back to the one-particle case described in the previous chapter. The same results are thus expected to follow, including the OTOC saturation, where the trajectory loops play a crucial role. Unlike the one-particle case, for the bosonic $N$-particle system on a lattice it is difficult to confront the theory to numerical simulations of fully chaotic systems and probe the temperature effects.   

\subsection{Systems without classical analogue}
\label{sec:spinchains}
In strong contrast to the previous examples of few-body systems with classical counterpart, there is the case interacting spin chains, which are many-body systems without a classical analogue. We describe below the case of chains of spins $1/2$ with a parameter in the Hamiltonian that can be used to tune the amount of chaos, as measured for instance, by level spacing statistics \cite{bohigas,poilblanc1993} or the spectral gap ratios \cite{Ata13,atas2013joint}. For simplicity we set in this section $\hbar=1$, and we call $L$ the total number of sites. The spin operators are 
\begin{equation}
    \hat{S}_i^{\mu}=\frac{1}{2}\hat{\sigma}_i^{\mu},
\end{equation}
where $i=0,1,\ldots,L-1$ labels the sites and $\hat{\sigma}^\mu$ are the Pauli operators for directions $\mu=x,y,z$. We use open boundary conditions, and since the spin operators are both unitary and Hermitian, the OTOC for infinite temperature can be written as
\begin{equation}
    \begin{aligned}
\mcC_{\hat{\sigma}^\mu_0 \hat{\sigma}^\nu_l}(l, t) &=-\frac{1}{2}\left\langle\left[\hat{\sigma}_{0}^{\mu}(t), \hat{\sigma}_{l}^{\nu}\right]^{2}\right\rangle \\
&=1-\operatorname{Re}\left\{\operatorname{Tr}\left[\hat{\sigma}_{0}^{\mu}(t) \hat{\sigma}_{l}^{\nu} \hat{\sigma}_{0}^{\mu}(t) \hat{\sigma}_{l}^{\nu}\right]\right\} / D,
\end{aligned}
\end{equation}
where  $D$ is the dimension of Hilbert space, and we have taken $\hW\equiv \hat{\sigma}_0^\mu$, and  $\hV\equiv \hat{\sigma}_l^\mu$. Where the factor $1/2$ was added to normalize the long-time limit to 1.

The Heisenberg chain with a random field in the $z$ direction is given by the Hamiltonian 
\begin{equation}
\label{eq:heisen}
    \hat{H}=\sum_{i=0}^{L-2}\left(\hat{S}_{i}^{x} \hat{S}_{i+1}^{x}+\hat{S}_{i}^{y} \hat{S}_{i+1}^{y}+\hat{S}_{i}^{z} \hat{S}_{i+1}^{z}\right)+\sum_{i=0}^{L-1} h_{i} \hat{S}_{i}^{z} \, ,
\end{equation}
where $h_i$ are independent random variables defined at each site, uniformly distributed in the interval $[-h,h]$. This model has been extensively used in studies of the many-body localization (MBL) transition (see Ref.~\onlinecite{alet2018many} for a review).
\vskip 0.25cm 

\begin{figure}
    \centering
    \includegraphics[width=0.95\linewidth]{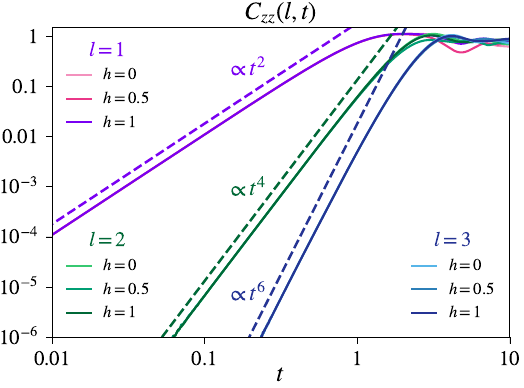}  %%{figs/spin_chain_short_time.pdf}  %%
    \caption{ Short-time behavior of the OTOC for Heisenberg chain with random field, with a length $L=9$ and fixed number $N=5$  of  spins up (or down), for spin separations $l=1,2,3$ (solid lines). The dashed lines correspond to the power law behavior obtained in Eq.~(\ref{eq:shortspin}). (Reproduced from Ref.~\cite{FortesPRE2019}, copyright 2019, American Physical Society.)}
    \label{Figtime}
\end{figure}

Plugging Eq.~\eqref{eqBCH} into the definition of 
$ {\cal C}_{\hat{\sigma}^\mu_0\hat{\sigma}^\nu_l}(t)$ yields the short-time behavior  \cite{Motrunich2018,Riddel2019,FortesPRE2019,craps2020lyapunov}
\begin{equation}
\label{eq:shortspin}
    {\cal C}_{\hat{\sigma}^z_0\hat{\sigma}^z_l} (t) \approx \frac{1}{2} \frac{t^{2 l}}{(l !)^{2}}
\end{equation}
for $l\ge 1$. Such a behavior is depicted in Fig.~\ref{Figtime}. Similar results, i.e. power law behavior for short times, can also be analytically obtained for other examples of spin chains like the perturbed XXZ model and the Ising model with tilted magnetic field \cite{FortesPRE2019}.

We note that the case of spin chains with transverse field was studied in Ref.~\onlinecite{craps2020lyapunov} for spins larger than $1/2$. In particular it is shown that in the large-spin limit, an exponential growth of the OTOC can be identified, consistently with the fact that in this limit, a classical analog system can be identified, and a Lyapunov exponent can be computed.  

%%%%%%%%%%
%%%%%%%%%%%%%%%%%%%%%%%
\subsection{Quantum chaos indicators from the long-time behavior of the OTOC}
\label{sec:QCIFLTB}
%%%%%%%%%%%%
The long-time behavior of the OTOC, in particular it saturation (mean) value  and the way it scales with the system size has been used to characterize chaos, or lack thereof. However, another useful approach is to use the knowledge extracted from the OTOC oscillations to produce an indicator (i.e. a measure) of quantum chaos.  

The study of Ref.~\onlinecite{FortesPRE2019} is focused on the fluctuations of the OTOC for large times in very different physical systems. The results for the kicked Harper map reproduced in Fig.~\ref{harperlongtime}, are qualitatively similar to the others where a transition from integrability to chaos is monitored, i.e.  while for integrable dynamics the OTOC presents large amplitude, in the strongly chaotic case, the corresponding curve is rather flat (with from small and random-like fluctuations). Taking this observation into account two chaos quantifiers were proposed. One measures localization in Fourier space, as the participation number 
\begin{equation}
\label{eq:xi}
\xi_{_\text{OTOC}}=\left[\int_{0}^{\infty} d \omega|\widetilde{\mcC}_{\hV\hW}(\omega)|^{4}\right]^{-1}   ,
\end{equation}
where $\widetilde{C}_{\hV\hW}(\omega)$ is the Fourier transform of $C_{\hV\hW}(t)$ on some time window $[t_{l},t_l+\Delta t]$ with $t_l\gg t^*$.  

The other quantifier 
\begin{equation}
\label{eq:sigma}
    \sigma_{_\text{OTOC}}=\sqrt{\langle \left( \mcC_{\hV\hW}(t) \right)^2 \rangle -\langle \mcC_{\hV\hW}(t)\rangle^2}
\end{equation}
measures the variance of the OTOC for the same fixed time-window as the one above-defined (a similar measure is studied in \cite{He2017}). The two previous OTOC-based measures exhibit a good agreement when compared with other indicators normally used to gauge the degree of chaoticity, like the Brody parameter $\beta $ \cite{BrodyRMP} (based on the eigenenergy spacing) or the inverse participation ratio $\xi_E$ (based on the eigenfunction distribution) \cite{FortesPRE2019,Borgonovi2019}.

\begin{figure}
    \centering
    \includegraphics[width=0.95\linewidth]{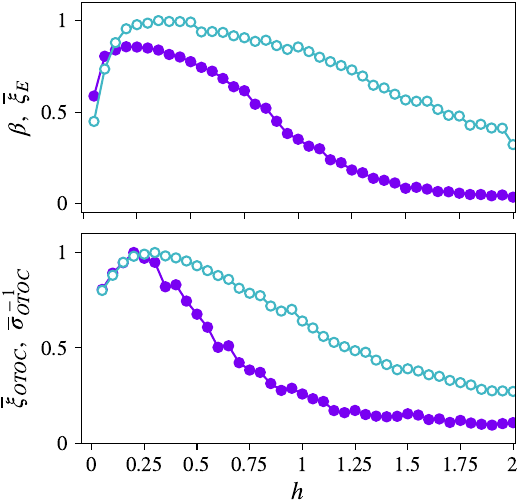} %%%{figs/figHeisenbergRandom.pdf}  
    \caption{ Chaos transition in the Heisenberg spin chain with a random magnetic field of amplitude $h$ quantified by different indicators. In the top panel two standard indications are shown: the inverse participation ration $\bar{\xi}_{E}$ (filled circles) and Brody parameter $\beta$ (open circles), averaged over 100 realizations, for a chain of length $L=13$, $N=5$ ($D=1287$). In the bottom panel two OTOC-based indicators are shown: $\bar{\xi}_{\text {OTOC}}$ (filled circles) and $\bar{\sigma}_{\text {OTOC}}^{-1}$ (open circles), averaged over 100 realizations for a chain of length $L=9$ and $N=5$ ($D=126$). (Reproduced from Ref.~\cite{FortesPRE2019}, copyright 2019, American Physical Society.)}
    \label{FigHeisen}
\end{figure}

In Fig.~\ref{FigHeisen} an example of OTOC-based indicators (\ref{eq:sigma}) and (\ref{eq:xi}) is shown (bottom panel) for the Heisenberg chain with random field defined by Eq.~(\ref{eq:heisen}), together with two standard chaos measures (bottom), the Brody parameter $\beta$ and the inverse participation ratio $\bar{\xi}_{E}$. A direct qualitative accord with both spectral measures can be observed. Additional evidence for other spin chains as well as for quantum maps are presented in  Ref.~\onlinecite{FortesPRE2019}. In the case of quantum maps the accuracy of the proposed measures can be put in evidence. For kicked maps, like the standard map or the Harper map, there are relatively large kicking strengths, deep into what is considered the chaotic regime, where small regular islands can appear in phase-space. Remarkably, the presence of this islands is systematically detected by both measures $\xi_{_\text{OTOC}}$ and $ \sigma_{_\text{OTOC}}$ as dips in the corresponding graphs (see \cite{FortesPRE2019}).
%%%%%%%%%%%%%%
%%%%%%%%%%%%%%

The importance of these measures is twofold. On the one hand they provide quantitative proof of the information about chaos that can be extracted from the long-time behavior of the OTOC. On the other hand, and as a consequence they allow to characterize the transition to chaos. In addition, these measures work even for very small system sizes, compatible with experimental setups readily available \cite{Li2017}. In \cite{fortes2020signatures} it is shown that the transition from chaos to integrable is qualitatively well described for spin chains of as short as $L=4$.

In Ref.~\onlinecite{Anand_coherence2021}, the authors show that the coherence generating power, which measures the average coherence generated under some time evolution, is deeply related the OTOC, and they use its variance at long times to characterize the transition to chaos.

%%%%%%%%%%%%%%%%%%%%%%
\subsection{Saddles}
%%%%%%%%%%%%%%
\label{sec:saddle}
In the fully chaotic case, hyperbolic periodic orbits, together with their associated stable and unstable manifolds, structure the classical dynamics, setting the global stability (i.e. the robustness against small perturbations) and the exponential divergence of nearby trajectories. In integrable or mixed systems, unstable equilibrium points in phase-space, and their associated separatrix, locally play a similar role, structuring the dynamics, and in their neighbourhood, yielding an exponential separation of trajectories along the separatrix. Such an exponential divergence in time has been pointed as a possible source of exponential growth of the OTOC in integrable systems \cite{hashimoto2020exponential,chavez2019quantum,PilatowskiPRE2020}. 

\begin{figure}
    \centering
    \includegraphics[width=0.95\linewidth]{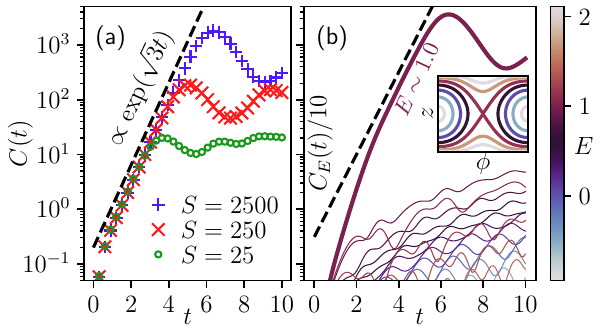}  %%{figs/fig_saddle1.png}  %%
        \caption{ (a) Exponential growth of the infinite-temperature OTOC of the integrable Lipkin-Meshkov-Glick model in the semiclassical limit. The growth saturates at the Ehrenfest time. The OTOC growth-rate $\Lambda_{\text {OTOC }}=\sqrt{3}$ agrees with the unstable exponent of the saddle point in the classical phase space. (b) Microcanonical-ensemble OTOCs  
    on an energy-window of 125 eigenstates with an average energy $E$. A few representative ensembles across the entire energy spectrum are shown. The one with $E \approx 1$, corresponding to the classical saddle, dominates the exponential growth observed in (a). (Inset) Energy landscape of the classical limit, with the same color code as (b) and the saddle in the center. (Reproduced from Ref.~\cite{XuPRL2020}, copyright 2020, American Physical Society.) }
    \label{Figsaddle1}
\end{figure}

An inverted harmonic oscillator is a paradigmatic case, characterized by a local unstable equilibrium point. However, the potential, not being bounded from below, prevents the introduction of temperature in any meaningful way. The addition of a quartic potential term to the one-dimensional inverted oscillator, leading to a double-well potential, has been used to calculate the OTOC in a one-particle model, and an exponential short-time growth of OTOC has been numerically obtained, in a limited time and temperature window, when the initial state is localized near the the separatrix \cite{hashimoto2020exponential}. The states in a narrow interval of energy close to that of the potential maximum have been shown to be associated with the observed exponential growth. Similar results have been observed when the OTOC is calculated in the regular regime of the Dicke \cite{chavez2019quantum,PilatowskiPRE2020} or Lipkin-Meshkov-Glick \cite{XuPRL2020} models, the two-site Bose-Hubbard model \cite{kidd2021saddle}, as well as in integrable many-particle bosonic systems near a quantum critical point \cite{HummelPRL2019} or the critical point of a dynamical phase transition in a spin-chain with long-range interactions \cite{Fazio2018}. 
In Fig.~\ref{Figsaddle1} the exponential growth is shown for the integrable Lipkin-Meshkov-Glick model as studied in \cite{XuPRL2020} for both infinite temperature and microcanonical ensemble.

%%%%%%%%%%%%%%%%%%
%%%%%%%%%%%%%%

We remark that the short-time exponential OTOC growth is less generic for systems presenting local maxima than in the fully chaotic case, since in the first scenario, it is linked to a fine-tuning of parameters. In particular, the temperature cannot be chosen lower than threshold given by the difference between the maximum and minimum of the potential. Moreover, as previously discussed (see \ref{sec:QCIFLTB}), the chaotic nature of a dynamical system is not only signing the short-time behavior, but also the long-time properties. Numerical studies of the Dicke model and a driven Bose-Hubbard dimer show that the short-time exponential growth associated to a saddle and that driven by chaos, can be efficiently distinguished by contrasting the long time-behavior in both cases, with strong oscillations characterizing the non-chaotic regime \cite{kidd2021saddle}. 
%%%%%%%%%%%%%%%%%%%%%%%%%%%%%%%%%%%
\section{Quantum bound to chaos through OTOC}
%%%%%%%%%%%%
In addition to bringing new life to
Larkin and Ovchinnikov's original idea of relating OTOC growth to chaos or quantum complexity
the 2016 article by Maldacena {\em et al.}  \cite{maldacena2016}  conjectured that the  growth of the OTOC has a bound which is given by the temperature of the system. That is, the exponential growth of the OTOC $\Lambda$ satisfies that,
\begin{equation}
    \Lambda \le 2 \pi k_{\rm B} T/\hbar,
    \label{bound}
\end{equation}
where $k_B$ is the Boltzmann constant and $T$ is the temperature of the system. This bound is believed to be completely universal and a fundamental property of quantum mechanics \cite{turiaci2019inelastic,jahnke2019chaos,kundu2022subleading}.
Moreover, its existence has a very special interest for theoretical physics. The fact that this bound is saturated by the 
Sachdev–Ye–Kitaev (SYK) model has been conjectured via its holographic duality to a realization of a  kind of black hole \cite{kitaev2015simple,MaldStanf2016,GuQiSanfort2017}, laying down a bridge between gravity, quantum mechanics, chaos and many-body. The SKY model \cite{Sachdev1993,rosenhaus2019introduction}, describing on-site randomly coupled large-$S$ spins or fermions with a two-body random interaction, provides a mean-field representation of the strange metal characterizing a high-temperature superconductor above its critical temperature, where no quasi-particle exists. The all-to-all random coupling of the SYK model is of the same nature of the two-body random interaction of the TBRE \cite{French1970,Bohiga1971,BrodyRMP} mentioned in the introduction. In models with a critical Fermi surface without quasi-particle excitations, the OTOC growth-rate and the butterfly velocity can be evaluated diagrammatically \cite{sachdev2017}, and the former can be shown to be well within the bound \eqref{bound}.

An important aspect to take into account regarding this saturation is that it was conjectured in the regularized version of the OTOC Eq. \ref{otocreg}.
The difference between regularized and non-regularized OTOC was first studied numerically  in interacting disordered metals
\cite{liao2018nonlinear}. Although both correlators have an exponential short time growth regime, the non-regularized one does not fulfill the bound given by Eq. \ref{bound}. This important aspect was further considered in Ref. \cite{tsuji2018bound} where a one-parameter family of out-of-time-ordered correlators is introduced 
showing that if all the elements of the family have an exponential growth, then the growth rate $\Lambda$ does not depend on the regularization parameter and satisfies the inequality \ref{bound}.

More recently, new connections to the OTOCs bound were found. In Refs. \cite{tsuji2018out,pappalardi2022quantum} it is shown that there is a profound relation between OTOCs and the
quantum fluctuation-dissipation theorem resulting in a physical mechanism that explain the bound of Eq. \ref{bound}. The fluctuation-dissipation theorem interrelates in a universal way the spectral characteristics of the fluctuations and the linear response of the same observable. Apparently these relationships
establish a smearing of fine time-details of correlations on a time scale $t=\hbar/\pi k_{\rm B} T$ that impose the bound. 
%%%%%%%%%%%%%%%%%%%%%%%%%%%%%%%%%%%%%%
\section{OTOCs in other contexts}
%%%%%%%%%%%%%%%%%%%
In addition to the topics previously discussed, the OTOC has been thoroughly studied in other interconnected centers of interest, briefly presented below. 
%%%
\subsection{Thermalization}
The unitarity of the quantum evolution presents a problem to the understanding of the ubiquitous phenomenon of thermalization. In the case of classical systems, such a process is understood assuming the chaotic behavior of systems with many degrees of freedom and a proper coarse-graining in phase space. In the case of isolated quantum many-body systems, there is a growing consensus that the foundations of their thermalization lay the so-called eigenstate thermalization hypothesis (ETH) \cite{srednicki1994chaos,deutsch1991quantum}. This hypothesis states that in a nonintegrable (quantum chaotic) Hamiltonian system, the energy eigenfunctions  correspond to a superposition of random waves, and the distribution of the eigenenergies is well described by random matrix theory, and thus the thermal properties of the system are embedded in each eigenstate.

An important connection between OTOCs and ETH has been established in Ref.~\onlinecite{murthy2019bounds}, showing that the bound of the OTOC is a consequence of the particular structure of the matrix elements of a few-body observable imposed by ETH \cite{deutsch1991quantum,srednicki1994chaos}. 
However, recent work \cite{foini2019eigenstate,chan2019eigenstate,brenes2021out} has shown that the initial exponential growth of the OTOCs for chaotic systems is in contradiction with the ETH assumptions implying the lack of correlation between matrix elements of local observables. Thus, the phenomenon of scrambling, or the short-time exponential growth, implies more structure than that required by ETH. 
%%%%
\subsection{Many-body localization}
%%%
The study of isolated interacting many-body systems in the presence of disorder has become very important in recent years \cite{basko2006,alet2018many}. 
These systems may not thermalize, because there is a memory of its initial condition in local observables due to the fact that energy eigenstates do not obey ETH and exhibit area law entanglement entropy \cite{pal2010many,abanin2017recent}. This phenomenon is referred to as many-body localization (MBL). 

While disorder has been shown in general to slow the onset of scrambling \cite{Swingle2017}, in the case of a MBL state such an effect can be extreme and the OTOC growth partially halted. The different behavior of the OTOC in systems that thermalize from those exhibiting MBL is remarkable. While in ergodic systems, the ballistic propagation of information is reveled as a linear propagation cone in the OTOCs dynamics, in MBL systems, 
early evidence for Heisenberg spin chains \cite{Chiara2006,BurrelLogLightcone2007} and disordered graphs \cite{Keating2007}, and further demonstrations in other many-body systems \cite{Kim2014,huang2017,He2017,Swingle2017,fan2017out,chen2017out} have shown that the propagation cone is logarithmic.

This qualitative feature of the OTOC dynamics has been used as a signature to detect MBL. For example, it has been used to characterize an intriguing intermediate dynamical phase in the interacting Aubry-Andr\'e model, appearing between the thermal and the many-body localized phases when the incommensurate potential strength is varied \cite{DasSarma2019}. Notwithstanding, logarithmic spreading of the OTOC has also been obtained in models without MBL \cite{smith2019logarithmic}.

\subsection{Open systems}
Quantum systems are in general not isolated, and the definition of a temperature-dependent OTOC presumes the contact of the system with a thermal environment. In this context, it is important to understand how information is scrambled and its relationship with the phenomena of dissipation and decoherence \cite{Galitski2018open}. This is particularly important in experiments when systems cannot be completely isolated or when imperfect reversal evolution is practically carried on. This last point was discussed in Ref. \cite{SwingleYunger2018} showing  that a proper renormalization can extract the ideal OTOC from imperfect experimental measurements.

The interplay between scrambling and dissipation was also considered in Ref.~\cite{Zhang2019} for the case of a spin chain coupled to an effective environment. It was shown that there is not only a leakage of information when an environment is present, but also a change in structure of the light cone is observed. 

The study of information scrambling from a quantum thermodynamics perspective has recently been undertaken  \cite{campisi2017,yunger2019Jarzynski,TouilPRX2021}. The 4-point out-of-time-order function ${\cal F}_{\hW\hV}$ has been related to a 2-point measurement scheme of an observable intertwined with an intermediate sudden perturbation \cite{campisi2017}. In particular, choosing the Hamiltonian as the observable, allowed to express the OTOC in terms of the second-moment of the work probability distribution. A Jarzynski-like equality has been derived for the OTOC, relating ${\cal F}_{\hW\hV}$ to an average over a work probability distribution set by the selected observables \cite{yunger2019Jarzynski}. Ref.~\onlinecite{TouilPRX2021} has shown that, using the mutual information as a scrambling quantifier in open quantum systems, it is possible to separate, in the entropy production of a non-unitary evolution, the contributions arising from decoherence and  scrambling.   

\subsection{Quantum information}
Entanglement is a basic property of quantum states that can be extended to temporal evolution \cite{zanardi2001entanglement}, while the OTOC characterizes the operator time-evolution. The link between them is therefore not evident, and indeed, it has been found in long-range spin chains that entanglement (quantified through the tripartite mutual information) and scrambling (characterized by the OTOC) have a rather different behavior in the chaotic, as well as in the integrable regimes \cite{Fazio2018}. However, defining a bipartite OTOC, with two unitary random operators $\hW_A$ and $\hV_B$ supported on two regions $A$ and $B$ of a bipartition and averaged according to the Haar measure, allows to connect ${\cal F}_{\hW_A\hV_B}$ with the quantum evolution of entanglement  \cite{styliaris2021information}. The relation between the entanglement operator dynamics and scrambling allows a  obtain rigorous results about entangling power, entropy productions and ${\cal F}_{\hW\hV}$. This results was initially obtained for infinite temperature limit of the initial state \cite{styliaris2021information} but then its was extended for generic thermal states in the case of regularized OTOCs \cite{anand2021brotocs}.

Mutual information can be used to measure entanglement. In Ref.~\onlinecite{Touil_2020} another interesting relation between entanglement and scrambling is shown: the growth of entanglement, quantified by the mutual information, is lower-bounded by the
time-dependent change of ${\cal F}_{\hW\hV}(t)$. 

The incompatibility of operators, one of the cornerstones of quantum mechanics, can be measured with uncertainty relations or by scrambling through OTOCs. Both measures are connected in Ref. \cite{yunger2019entropic}, arriving to an entropic uncertainty relations for scrambling.

\section{OTOC and EXPERIMENTS}
\noindent

In recent years, OTOC measurements have  been realized in several experimental setups.  Systems of different nature have been excellent platforms to measure the scrambling of quantum information. These systems include trapped ions \cite{Garttner2017,landsman2019verified, mi2021information}, nuclear magnetic resonance (NMR) \cite{Li2017,Xinfang2020,MohamadPRR2020,SanchezPEL2020},  or superconducting qubits \cite{braumuller2022probing}. Although the experimental results are very promising, the number of particles in the many-body systems is still small.  This is due to the extreme degree of control of the evolution must have, especially to generate the time reversal necessary to compute OTOCS. This is tricky in highly entangled, many-body systems.

Several important aspects of many-body dynamics have been scrutinized in labs through OTOCs. For example, in a quantum simulator of an Ising spin chain with more that 100 ions in a Penning trap, the measurement of the OTOC gives information of the state of the system and the way that different parts of the system are correlated \cite{Garttner2017}. An ion-trap setup with seven qubit circuit was used to measure scrambling of an unitary process of conditional teleportation that involves only three of them \cite{landsman2019verified}. 
OTOCs has also been measured in a 87 Rb Bose-Einstein condensate
in which the paradigmatic Kicked top model is simulated 
 mapping the angular momentum projection states of an effective quantum spin onto the linear momentum states of the condensate
 \cite{Meier2019}.

NMR is an experimental platform very suitable for OTOC experiments due to its versatility to perform backwards evolution. 
It was shown on a NMR Quantum Simulator consisting of 4 spins \cite{Li2017}, that the OTOC has different long-time behavior depending on the chaotic or integrable evolution. In a similar system \cite{Xinfang2020},  equilibrium and dynamical quantum phase transitions from quench dynamics of OTOC were observed. In solid-state NMR, correlations between a central spin and the environment were studied using OTOCs \cite{MohamadPRR2020} and the connection between the perturbation independent decay of the Loschmidt echo and the retrieval of scrambled information \cite{SanchezPEL2020}.   

Other systems used to measure OTOCs are 
superconducting qubits. Quantum thermalization and information scrambling were studied in the $XX$-ladder model and the one-dimensional $XX$ model measuring OTOCs in a ladder-type superconducting quantum processor \cite{zhu2022observation}.
In Ref. \cite{braumuller2022probing}, a $3\times 3$ two-dimensional hard-core Bose-Hubbard lattice was implemented and signatures of 2-D many body localization was observed measuring OTOCs.

\section{CONCLUDING REMARKS}
\noindent
%%%
This article reviewed the deep connections between the OTOC and Quantum Chaos. It has been pointed out that the hasty conclusion of identifying an exponential short-time growth of the OTOC as a signature of chaos should be taken with a grain of salt. A similar precaution should also be adopted for the opposite case: non-exponential growth can be found in systems widely considered chaotic. Therefore, a refined analysis based on different time-scales of the OTOC has been performed, showing that important features of Quantum Chaos can be found in short, intermediate and long-time regimes, depending on the type of system and the chosen gauges. 

Exponential OTOC growth for short times is ascribed to a quantum Lyapunov regime. In the case where a classical Lyapunov regime exists for the underlying classical system, there is strong evidence (both analytical and numerical) supporting the correspondence between both regimes. This correspondence is non-trivial in the temperature-dependent case, where the effect of an energy-dependent Lyapunov exponent has to be taken into account. Similarly, for chaotic systems that are not uniformly hyperbolic, the phase-space average of different Lyapunov exponents makes the previous connection more involved. It is important to stress that for systems with a classical hyperbolic exponential instability, the above-discussed correspondence also holds, even if chaos is not present. Note however that in the OTOC computed in the latter case there is no thermal average performed and the initial state plays a prominent role.

An important point highlighted in this work concerns the behavior of the OTOC for long times, where a deep connection exists between the integrability of the dynamics with key OTOC features, like the average infinite-time value and the fluctuations around it. These OTOC quantities can be used to gauge, up to a very good accuracy, the amount of chaos or integrability present in the system. 

The intermediate-time regime has shown to also be of interest. This intermediate time-window around the scrambling time, when the OTOC growth ceases, characterizes the approach equilibrium for chaotic systems. In the latter case, as suggested in \cite{polchinski2015} and demonstrated in \cite{OTOC_gato_PRL}, the intermediate-time regime is dominated by the largest Ruelle-Pollicott resonance. This is a very important feature in the connection between OTOC and Quantum Chaos, as it is well-known that there are two main features characterizing classical chaotic systems: exponential separation of initial conditions (a.k.a butterfly effect) for relatively short times, and the mixing that follows, resulting of the folding-back of classical trajectories in a finite-size system. 

The case of quantum chaotic maps has been described in detail because they are very simple systems where the OTOC can be shown to exhibit Lyapunov and Ruelle features in the short and intermediate time regimes, respectively. It remains as an open problem the determination of the two previous features in the OTOC for more complex systems.

In many-body systems, where the classical limit is less obvious (if there is one), the measures of chaos defined from the OTOC for long times play an important role. They have been shown to efficiently identify the ergodic phase and the departures from it, and moreover, such gauging works even for system-sizes small enough to be experimentally accessible. This remarkable observation can have consequences in identifying the many-body localized phase, essentially defined as a non-ergodic phase where ETH fails \cite{alet2018many}, for the case of small system-sizes.

%%%%%%%%%%%%%%%%%%%%%%%%%%%%%%%%%%%%%%%%%%%%%%%%%%%%%%%%%%%%%%%%%%%%%%%%%%%%%%%%
%  \bibliographystyle{apsrev4-1}
%\bibliography{refs}

%%%%%%%%%%%%%%%%%%%%%%%%%%%%%%%%%%%%%%%%%%%%%%%%%%%%%%%%%%%%%%%%%%%%%%%%%%%%%%%%
\end{document}		%%%%  		***				FIN				***			%%%%
%%%%%%%%%%%%%%%%%%%%%%%%%%%%%%%%%%%%%%%%%%%%%%%%%%%%%%%%%%%%%%%%%%%%%%%%%%%%%%%%